\documentclass{article} 
\usepackage{iclr2026_conference,times}
\usepackage{graphicx}

\usepackage{amsmath,amsfonts,bm}

\def\eqref#1{equation~\ref{#1}}

\def\1{\bm{1}}

\DeclareMathAlphabet{\mathsfit}{\encodingdefault}{\sfdefault}{m}{sl}
\SetMathAlphabet{\mathsfit}{bold}{\encodingdefault}{\sfdefault}{bx}{n}

\usepackage{hyperref}
\usepackage{url}
\usepackage{wrapfig}
\usepackage{booktabs} 
\usepackage[table]{xcolor}
\usepackage{multirow}
\usepackage{amsfonts}
\usepackage{verbatim}
\usepackage{placeins}
\usepackage{bbm}
\usepackage{bm}

\usepackage{amsmath}
\usepackage{amssymb}
\usepackage{mathtools}
\usepackage{amsthm}

\title{NMIRacle: Multi-modal Generative Molecular Elucidation from IR and NMR Spectra}

\author{Federico Ottomano${}^1$, Yingzhen Li${}^2$, Alex M. Ganose${}^1$\\
${}^1$Department of Chemistry, Imperial College London\\
${}^2$Department of Computing, Imperial College London\\
\texttt{f.ottomano@imperial.ac.uk} \\
}

\iclrfinalcopy 
\begin{document}

\maketitle

\begin{abstract}
Molecular structure elucidation from spectroscopic data is a long-standing challenge in Chemistry, traditionally requiring expert interpretation. We introduce NMIRacle, a two-stage generative framework that builds upon recent paradigms in AI-driven spectroscopy with minimal assumptions. In the first stage, NMIRacle trains a generator to reconstruct molecular structures from count-aware fragment representations, capturing both fragment identities and their occurrences. In the second stage, a spectral encoder maps input spectra (IR, $^1$H-NMR, $^{13}$C-NMR) into a latent embedding used to condition the pre-trained generator, which is fine-tuned for direct spectra-to-molecule generation. This formulation bridges fragment-level chemical modeling with spectral evidence, yielding accurate molecular predictions. Empirical results demonstrate that NMIRacle outperforms existing baselines on molecular elucidation, while maintaining robust performance across increasing levels of molecular complexity. NMIRacle code is publicly available at NMIRacle code is publicly available at \href{https://github.com/fedeotto/nmiracle}{https://github.com/fedeotto/nmiracle}.
\end{abstract}

\section{Introduction}
\label{intro}
Determining the molecular structure of an unknown compound through spectroscopy is a fundamental problem in Chemistry, central to drug discovery, metabolomics, and materials design.
This task is challenging due to the combinatorial explosion of possible atomic arrangements: even for molecules with fewer than 36 heavy atoms, the size of drug-like chemical space could exceed $\sim 10^{33}$~\citep{polishchuk2013molsize}. Techniques including \textit{infrared} (IR) spectroscopy, \textit{nuclear magnetic resonance} (NMR) spectroscopy and \textit{mass spectrometry} (MS) provide complementary yet indirect evidence of the molecular structure, and interpreting them requires integrating heterogeneous and often noisy signals.
Traditionally, structure elucidation relies on expert-driven spectral interpretation or database matching.
These strategies are limited by subjectivity, the need for extensive chemical expertise, and the inability to identify molecules absent from reference libraries.
Recent advances in deep learning have opened new directions for automated elucidation, including (i) cross-modal retrieval systems that learn shared embeddings of spectra and molecular structures~\citep{yang2021crees, jin2025nmrsolver,mirza2024elucidating}, and (ii) \emph{de novo} generative frameworks that direcly predict molecular graphs or sequences from spectroscopic evidence~\citep{bohde2025diffms, litsa2021spec2mol, guo2024can, yang2026diffnmr}.
While retrieval-based methods leverage existing databases to identify the closest-matching structures, \emph{de novo} generative approaches do not depend on pre-existing molecular libraries, making them inherently more flexible and capable of proposing novel compounds. However, this fully generative formulation poses substantial challenges: the model must integrate multiple spectra modalities with distinct noise characteristics 
and resolution biases, and learn a high-dimensional, multimodal mapping from continuous spectra to discrete molecular representations. A more comprehensive discussion of related work in Appendix~\ref{app:related_work}.
Despite 
the availability of new datasets and benchmarks~\citep{bushuiev2024massspecgym, guo2024can}, current spectra-to-molecule generative methods typically exhibit one or more limitations: (i) reliance on a single spectral modality, which neglects complementary patterns~\citep{litsa2021spec2mol, bohde2025diffms, bushuiev2024massspecgym}; (ii) dependence on extensive pre-processing (e.g., peak extraction, multiplet assignment) to convert spectra into symbolic or text-based inputs~\citep{alberts2023learning, yao2023cmgnet, jin2025nmrsolver}; (iii) assumptions of strong prior information, such as chemical formula or molecular scaffold \newpage \citep{alberts2024unraveling, wang2025madgen}, which are rarely available under realistic experimental conditions; iv) limited benchmarking settings, restricted to molecules composed of only a few chemical species (typically C, N, O) and fewer than $20$ heavy (non-hydrogen) atoms~\citep{hu2024accurate}. \\
In this work, we tackle the most challenging formulation of molecular structure elucidation: direct generation of molecular structures from raw, multi-spectra input. We build upon previous established paradigms 
in data-driven molecular elucidation from spectroscopy with minimal assumptions~\citep{hu2024accurate}.
We introduce \textbf{NMIRacle}, a generative framework that learns from spectroscopic intensity arrays, the same data produced by experimental instruments, requiring only minimal pre-processing to handle modality-specific signal characteristics. 
This setup is intentionally difficult, as the model must infer structural constraints from noisy, high-dimensional inputs, but it enables greater realism and generalization across multiple acquisition settings. 
We evaluate NMIRacle on a multimodal spectroscopic dataset comprising molecules with up to 35 heavy atoms and diverse chemical compositions~\citep{alberts2024unraveling}. 
Our framework consistently obtains strong molecular elucidation performance across a broad range of molecular sizes and structural complexities. We summarize the main contributions of this work below:

\begin{itemize}
    \item We propose NMIRacle, a generative framework for molecular structure elucidation from spectroscopy, operating directly on combinations of raw IR, ${}^1$H-NMR, and ${}^{13}$C-NMR spectra.

    \item We leverage count-aware fragment representations as an alternative to the binary indicators commonly used in existing frameworks. We demonstrate that capturing fragment occurrences provides a more faithful structural representation of molecules that effectively transfers to the downstream spectra-to-molecule task.

    \item We design a multi-spectral encoder that fuses raw IR, ${}^1$H-NMR, and ${}^{13}$C-NMR signals through intra- and inter-spectral attention. 
    \item We demonstrate strong performance on molecular elucidation and robust generalization to complex molecules under minimal input assumptions.
\end{itemize}

\section{Methods}
\subsection{Problem formulation}
We formulate molecular structure elucidation as a conditional generative modeling task. 
Given a set of complementary spectroscopic measurements 
$\mathcal{S} = \{\mathbf{s}_1, \mathbf{s}_2, \dots, \mathbf{s}_N\}$, 
where each $\mathbf{s}_i \in \mathbb{R}^{n_i}$ is a raw intensity vector sampled over the measurement domain of modality $i$, 
the goal is to generate the corresponding molecular structure $\mathcal{M}$. While the number and type of modalities $N$ are fixed for a given model instance, our framework can be applied to any subset of available spectroscopic data. In practice, we represent each molecule by its SMILES sequence 
$\mathbf{y} = (y_1, y_2, \dots, y_T)$,
which provides full information about atom types and connectivity.
We assume access to a dataset $\mathcal{D} = \{(\mathcal{S}^{(m)}, \mathbf{y}^{(m)})\}_{m=1}^M$ of paired spectra–molecule examples. 
The learning objective is to estimate model parameters $\theta$ that maximize the likelihood of generating the correct SMILES given the corresponding input spectra:
\begin{equation}
	\theta^* = 
	\arg\max_{\theta}
	\mathbb{E}_{(\mathcal{S}, \mathbf{y}) \sim \mathcal{D}} 
	[\log p_\theta(\mathbf{y} \mid \mathcal{S})] \, .
\end{equation}
Each spectral modality provides complementary structural evidence. We focus on three common techniques: IR spectroscopy captures vibrational modes of molecular bonds; \emph{proton-NMR} (${}^1$H-NMR) spectroscopy measures hydrogen environments and connectivity; \emph{carbon-NMR} spectroscopy (${}^{13}$C-NMR) probes carbon backbone structure~\citep[Chapter~13]{clayden2012organic}. 
\subsection{Spectra pre-processing}
We convert raw spectral data from different analytical techniques into unified sequence representations suitable for transformer-based processing.
\paragraph{IR and ${}^1\text{H-NMR}$} We apply minimum amount of pre-processing for these spectra modalities, since peak shapes and relative intensities contain valuable structural information. These are normalized to the $[0,1]$ range to ensure consistent intensity scales across samples. These continuous intensity profiles are used directly as inputs to the model.
\paragraph{${}^{13}\text{C-NMR}$} For carbon-NMR, peak intensities are not reliable indicators of carbon counts and so, following previous work~\citep{mirza2024elucidating}, we focus on chemical shift positions rather than intensities. We detect peaks in the raw array using SciPy's \texttt{find\_peaks} function with a threshold of 10\% relative to the maximum intensity. The detected peak positions are mapped from array indices to chemical shift values across the 0–220 ppm range.
This range is then discretized into 80 equal-width bins of approximately $2.75$ ppm each, and we create a binary vector indicating the presence or absence of peaks in each bin.

\subsection{Method overview}
\begin{figure*}[!t]
\centering
\includegraphics[width=0.98\textwidth]{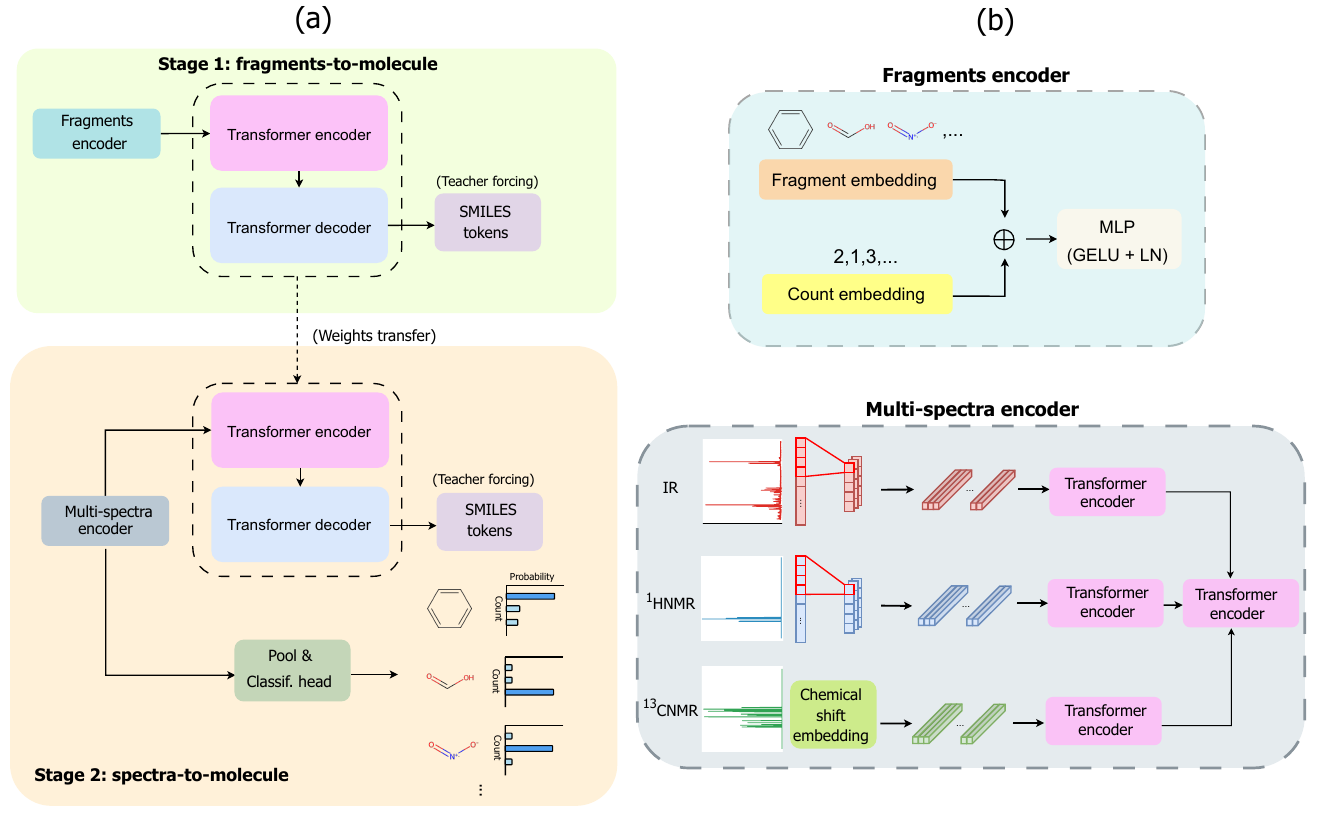}
    \caption{
    Overview of NMIRacle. 
    (a) The model is trained in two stages: 
    Stage~1 learns a fragment-conditioned molecular generator that reconstructs full molecular structures from count-aware fragment representations, establishing a molecular prior $p_\phi(\mathbf{y}\mid\mathbf{c})$. 
    Stage~2 introduces a multi-spectra encoder that maps combinations of raw IR, ${}^1$H-NMR, and ${}^{13}$C-NMR spectra into latent embeddings $\mathbf{z}_\psi(\mathcal{S})$, used to condition the pre-trained generator for direct spectra-to-molecule generation. 
    (b) The architecture integrates (i) a count-aware fragment encoder that embeds molecular fragments and their occurrences, and (ii) a multi-spectra encoder that fuses complementary spectra modalities into a unified latent representation.
    }
	\label{fig:model}
\end{figure*}

We conceptualize molecular structure generation from spectra as a two-stage conditional generative process, building upon previous work~\citep{hu2024accurate, bohde2025diffms}. Figure~\ref{fig:model} provides a visual overview of the proposed framework.

\paragraph{Global fragment vocabulary}
We define a vocabulary of chemical substructures 
$\mathcal{V} = \{ f_1, f_2, \dots, f_{|\mathcal{V}|} \}$ 
that serves as a discrete compositional basis for representing fragment compositions of molecules. Specifically, we curate a fragments vocabulary including 991 SMARTS patterns covering a broad range of common organic motifs.
\paragraph{Molecular representation}
We consider two representations for a molecule $\mathcal{M}$:  
(i) a fine-grained sequence of SMILES tokens 
$\mathbf{y} = (y_1, y_2, \dots, y_T)$ encoding full information about atom types and connectivity;  
(ii) a coarse fragments vector $\mathbf{c}$ capturing the fragment composition of the underlying molecule.  Specifically, $\mathbf{c} = (c_1, c_2, \dots, c_{|\mathcal{V}|}) \in \mathbb{N}_0^{|\mathcal{V}|}$,
where $c_j$ indicates the number of occurrences of fragment $f_j$ in $\mathcal{M}$.  
This compact representation captures information about both fragment identities and their occurrences.

\paragraph{Two-stage modeling}
The overall spectra-to-molecule model is trained in two stages: (i) In Stage~1, we pre-train a fragment-conditioned generative model
$p_\phi(\mathbf{y} \mid \mathbf{c})$ 
that learns to reconstruct a molecular SMILES sequence from its corresponding fragment composition $\mathbf{c}$; 
(ii) In Stage~2, a spectra encoder $q_\psi$, trained from scratch, maps spectroscopic measurements $\mathcal{S}$ into a continuous embedding 
$\mathbf{z}_\psi(\mathcal{S})$ that conditions the pre-trained generator from Stage~1. Under this formulation, the latter is fine-tuned to approximate the true conditional distribution
\begin{equation}
    p(\mathbf{y} \mid \mathcal{S}) 
    \approx p_\phi(\mathbf{y} \mid \mathbf{z}_\psi(\mathcal{S})) \, .
    \label{eq:determ_mapping}
\end{equation}
Conceptually, this can be viewed as replacing the marginalization
\begin{equation}
    p(\mathbf{y} \mid \mathcal{S}) 
    = \sum_{\mathbf{c}} p(\mathbf{y} \mid \mathbf{c}) \,
      q(\mathbf{c} \mid \mathcal{S}) \, ,
\end{equation}
with a deterministic point estimate of the fragment composition  induced by the spectral embedding $\mathbf{z}_\psi(\mathcal{S})$. In other words, $\mathbf{z}_\psi(\mathcal{S})$ serves as a continuous surrogate for the (unknown) fragment composition $\mathbf{c}$, enabling the generator to transfer from fragment-conditioned pre-training to spectra-conditioned fine-tuning.

\subsubsection{Stage 1: Fragments-to-molecule pre-training}
In the first stage, we learn parameters $\phi$ of a conditional generative model $p_\phi(\mathbf{y} \mid \mathbf{c})$ that reconstructs a molecular SMILES sequence 
$\mathbf{y} = (y_1, y_2, \dots, y_T)$ from a corresponding coarse fragments vector $\mathbf{c}$. 
Previous 
approaches typically adopt a binary fragment encoding, where each entry $c_j \in \{0,1\}$ indicates the presence or absence of fragment $f_j$~\citep{bohde2025diffms, hu2024accurate}.
In contrast, we employ a count-aware fragment representation, where $c_j \in \mathbb{N}$ denotes the number of occurrences of each fragment in the molecule.
This representation provides a more faithful description of molecular composition, enabling the model to capture structural regularities that depend on fragment repetition (e.g., ring patterns, chain extensions).
Each fragment type $f_j$ and its associated count $c_j$ are independently embedded:
\begin{equation}
    \mathbf{h}_{f_j} = \text{Embed}_f(f_j) , \quad
    \mathbf{h}_{c_j} = \text{Embed}_c(c_j) \in \mathbb{R}^{d} ,
\end{equation}
where $\text{Embed}_f(\cdot)$ and $\text{Embed}_c(\cdot)$ denote learnable lookup tables for fragment types and occurrences, respectively, while $d$ indicates the hidden dimensionality.
The two embeddings are combined through element-wise addition, followed by a non-linear transformation:
\begin{equation}
    \mathbf{h}_j = \text{LayerNorm}(\text{MLP}(\mathbf{h}_{f_j} + \mathbf{h}_{c_j})) \in \mathbb{R}^d \, ,
\end{equation}
where \text{MLP} denotes a single-hidden-layer perceptron with GELU activation.  
The resulting set of count-aware fragment embeddings $\{\mathbf{h}_j\}$ serves as input tokens to the transformer encoder, which provides contextualized representations for decoding. Conditioned on this context, the decoder autoregressively predicts SMILES tokens:
\begin{equation}
    \mathcal{L}_{\text{Stage1}}(\phi) = \mathbb{E}_{({\mathbf{c}, \mathbf{y}})}\Big[- \sum_{t=1}^{T} 
    \log p_\phi(y_t \mid y_{<t}, \{\mathbf{h}_j\})\Big] ,
    \label{eq:stage1_loss}
\end{equation}
minimizing the standard autoregressive negative log-likelihood.
\subsubsection{Stage 2: spectra-to-molecule fine-tuning}
In the second stage, we fine-tune the fragment-conditioned generator $p_{\phi}(\mathbf{y} \mid \mathbf{c})$, previously trained under the count-aware fragment encoding scheme,
to map spectroscopic measurements $\mathcal{S}$ directly to molecular SMILES. 
Rather than conditioning on count-aware fragment encodings $\{\mathbf{h}_j\}$, the model now conditions on 
latent spectral embeddings produced by a multi-spectral encoder $q_{\psi}$ (Eq.~\ref{eq:determ_mapping}). 
These embeddings serve as a continuous proxy for the fragment-level representation learned in pre-training, 
thereby preserving the same generative interface while adapting it to spectral inputs.

\paragraph{Multi-spectra encoder} Each input spectrum $\mathbf{s}_i \in \mathbb{R}^{n_i}$ from modality $i \in \{\text{IR},\,{}^1\text{H-NMR},\,{}^{13}\text{C-NMR}\}$ is processed by a modality-specific encoder $E_{\text{spec}}^{(i)}$. 
The encoder extracts spectral features and projects them into a shared embedding space of dimension $d$.
For IR and ${}^1\text{H-NMR}$ spectra, we first apply 1D convolutional layers to capture local peak patterns and compress the signal into feature maps $\mathbf{Z}_i \in \mathbb{R}^{s_i \times c_i}$. 
A learnable linear projection $P^{(i)} \in \mathbb{R}^{c_i \times d}$ maps these features to token embeddings with hidden dimensionality $d$. 
To retain spectral ordering, we add learnable positional encodings $\mathbf{W}^{pos}_i \in \mathbb{R}^{s_i \times d}$:

\begin{equation}
    \mathbf{Z}_i^{\text{seq}} = P^{(i)}(\mathbf{Z}_i) + \mathbf{W}^{pos}_i \, .
\end{equation}

We ablate the impact of learnable positional encodings against sinusoidal positional encodings in Appendix~\ref{app:ablations}. For $^{13}\text{C}$-NMR spectra, where inputs are discrete chemical shift indices rather than continuous peaks, 
we omit positional encodings and instead use a learnable embedding lookup for each non-zero bin index.
Each modality sequence $\mathbf{Z}_i^{\text{seq}}$ is then passed to an intra-modal transformer encoder to model local dependencies among peaks within the same spectrum
\smallskip
\begin{equation}
    \mathbf{H}_i = \text{TEnc}_{\text{intra}}^{(i)}(\mathbf{Z}_i^{\text{seq}}) \in \mathbb{R}^{s_i \times d} ,
\end{equation}
producing modality-specific embeddings. 
The encoded modalities are concatenated and fed to a separate, inter-modal transformer encoder:
\smallskip
\begin{equation}
    \mathbf{H}_{\text{inter}} = \text{TEnc}_{\text{inter}}([\mathbf{H}_1 ;
    \mathbf{H}_2 ; \mathbf{H}_3]) \in \mathbb{R}^{s \times d} \, ,
\end{equation}
where $;$ denotes concatenation along the sequence dimension, and $s = \sum_i s_i$ is the resulting sequence length.
This 
enables dedicated learning between distinct modalities (e.g., associating IR absorption bands with ${}^1$H chemical shifts linked to the same functional groups).
The obtained representation $\mathbf{H}_{\text{inter}}$ replaces the count-aware fragment tokens $\{\mathbf{h}_j\}$ 
used in Stage~1 (Eq.~\ref{eq:stage1_loss}) as contextual input to the pre-trained model $p_\phi$, 
thus conditioning molecular generation directly on spectral features.
\paragraph{Fragment composition head}
To enhance fragment-level supervision, we adopt a multi-task setup~\citep{hu2024accurate} optimizing concurrently the model for SMILES reconstruction (Eq. \ref{eq:stage1_loss}) and for predicting fragment compositions. 
First, the fused representation $\mathbf{H}_{\text{inter}}$ is mean-pooled to a global feature vector:
\begin{equation}
    \mathbf{h}_{\text{inter}} = \text{MeanPool}(\mathbf{H}_{\text{inter}}) \in \mathbb{R}^d .
\end{equation}
Each fragment identity $f_j$ is represented by a one-hot vector $\mathbf{e}_{f_j}$ from the fragment vocabulary. For each fragment, we concatenate $\mathbf{h}_{\text{inter}}$ and $\mathbf{e}_{f_j}$ 
and predict a categorical distribution over possible counts:
\begin{equation}\label{eq:fragment_head}
    p_\psi(c_j \mid \mathcal{S}, f_j) = 
    \text{Softmax}\!\left( \text{MLP}[\mathbf{h}_{\text{inter}} ; \mathbf{e}_{f_j}] \right),
\end{equation}
where $c_j \in \{c_0, \dots, c_{\text{max}}\}$, $c_{\text{max}}$ represents the maximum observed occurrences of a fragment in a molecule, $\text{MLP}(\cdot)$ denotes a single-hidden-layer perceptron with GELU activation, and $;$ denotes feature-wise concatenation. This formulation enables the model to learn both fragment presence and occurrence directly from spectral evidence.

\paragraph{Training objective}
During Stage~2, the pre-trained generator $p_\phi(\mathbf{y} \mid \mathbf{c})$ from Stage~1 is fine-tuned with spectra conditioning, 
while the spectra encoder $q_\psi$ is trained from scratch.
The overall objective combines (i) a sequence-level cross-entropy loss for molecular reconstruction and (ii) a fragment-level cross-entropy loss over discrete fragment occurrences:
\begin{align}
\mathcal{L}_{\text{Stage2}}&(\phi, \psi)
= \alpha \, \mathbb{E}_{(\mathcal{S}, \mathbf{y})}
\Big[ - \sum_{t=1}^{T} \log p_\phi(y_t \mid y_{<t}, \mathbf{z}_\psi(\mathcal{S})) \Big] \nonumber \\
&\quad + \beta \, \mathbb{E}_{(\mathcal{S}, \mathbf{c})}
\Big[ - \sum_{j=1}^{|\mathcal{V}|} \log p_\psi(c_j \mid \mathcal{S}, f_j) \Big] \, ,
\label{eq:stage2_loss}
\end{align}
where $\alpha$ and $\beta$ balance the contributions of the two tasks, $p_{\phi, \psi}$ denotes the fine-tuned generator with spectra conditioning, and $p_\psi(c_j \mid \mathcal{S}, f_j)$ parameterizes the fragment composition head (Eq.~\ref{eq:fragment_head}). In practice, we set $\alpha = \beta=1$.
This multi-task setup encourages the latent representation $\mathbf{z}_\psi(\mathcal{S})$ to encode fragment compositions for molecular generation.
\section{Experiments}
\subsection{Datasets}
We employ two complementary datasets for our experiments: a molecular pre-training dataset for Stage 1 and a spectra fine-tuning dataset for Stage 2.
\paragraph{Molecular pre-training dataset} 
We build upon an existing molecular dataset employed in previous work~\citep{hu2024accurate}, comprising approximately $\sim 3.1$M molecules, obtained by combining $\sim3$M compounds randomly sampled from the GDB-17 database~\citep{ruddigkeit2012enumeration} with an additional $\sim140$k entries sourced from SpectraBase~\citep{spectrabase}.
While this collection provides a large set of molecules, it is chemically-limited, containing only carbon (C), oxygen (O), and nitrogen (N) atoms, and restricted to a maximum of $19$ heavy atoms per molecule.
\begin{wrapfigure}[26]{r}{0.47\textwidth}
\vspace{-10pt}
\centering
\includegraphics[width=0.48\textwidth]{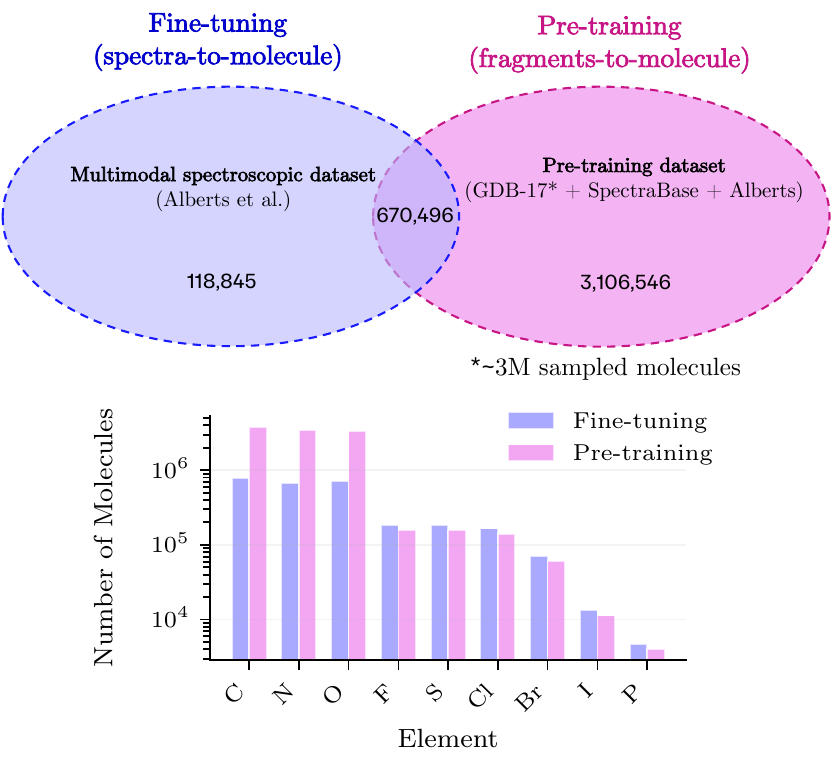}
\caption{
(Top) Venn diagrams illustrate the overlap between the molecular pre-training dataset (derived from GDB-17 and SpectraBase) and the additional molecules incorporated from~\citet{alberts2024unraveling} dataset.
(Bottom) Element distribution across the utilized datasets, highlighting the broader chemical diversity introduced by the data augmentation.
}
\label{fig:datasets}
\vspace{-10pt}
\end{wrapfigure}
Such constraints make it poorly representative of the molecular diversity encountered in experimental settings. 
To address this limitation, we extend the original pre-training pool with $\sim670$k molecules from a recent multimodal spectroscopic dataset introduced by~\citet{alberts2024unraveling}. 
This augmentation increases chemical diversity up to 9 distinct elements and extends molecular size up to $35$ heavy (non-hydrogen) atoms, thereby exposing the pre-training model to richer compositional and structural variations. 
\paragraph{Spectra fine-tuning dataset} 
We utilize a recently proposed multimodal spectroscopic dataset~\citep{alberts2024unraveling} as the main benchmark for spectra-to-molecule task (Stage 2). It contains over $\sim790$k molecules paired with various simulated spectra, including IR, ${}^1$HNMR and $^{13}$CNMR. We split the dataset into training, validation and test subsets in an 8:1:1 ratio. The training split ($\sim 670$k SMILES) corresponds to the augmentation performed on the molecular pre-training dataset. Crucially, we ensure no molecules from either pre-training or training data are present in the test set. This guarantees that the final evaluation measures the model’s ability to predict entirely unseen molecules, only from spectral evidence.

\subsection{Baselines}
We employ different baselines to compare the performance of the proposed approach.

\paragraph{SMILES/SELFIES transformers} We implement transformer models that operate on simple concatenations of spectra features. To stay consistent with the proposed methodology, which assumes minimal pre-processing on input spectra, we apply minimal feature extractors: 1D convolutional layers for continuous spectra (IR, ${}^1$H-NMR) and a learnable lookup embedding for ${}^{13}$C-NMR peak bins. 
The resulting representations are concatenated across modalities and provided to an encoder-decoder transformer that generates molecules in either SMILES or SELFIES format.  
This setup is inspired by the benchmark provided in the work of \citet{alberts2024unraveling}, but differs in that we avoid domain-specific pre-processing (e.g., MestreNova~\citep{mestrelab_mnova} peak extraction) and instead let the neural encoders discover spectral patterns directly from raw data.
\paragraph{NMR2Struct}
We evaluate NMR2Struct~\citep{hu2024accurate}, a two-stage framework for spectra-to-molecule prediction. In Stage~1, a Transformer-based molecular generator is pre-trained to reconstruct SMILES sequences from binary fragment indicators.
In Stage~2, a spectral encoder is trained to produce embeddings that condition the pre-trained generator, which is fine-tuned for spectra-to-molecule generation. We re-implement NMR2Struct within our experimental setting to ensure a controlled and fair comparison.
\paragraph{Spec2Mol}
We evaluate Spec2Mol~\citep{litsa2021spec2mol}, which follows a two-stage, latent-space alignment approach. In Stage~1, a GRU-based SMILES autoencoder is trained to reconstruct molecular SMILES, yielding a continuous latent representation.
In Stage~2, a convolutional spectra encoder is trained to minimize the $\ell_2$-loss between spectral embeddings and SMILES latent representations produced by the autoencoder encoder.
At inference, spectra are embedded using the trained spectra encoder and decoded into SMILES using the autoencoder decoder.
We re-implement Spec2Mol within our experimental setting to ensure a controlled and fair comparison.
\subsection{Results and discussion}
Table~\ref{tab:sub2struct} reports results for the pre-training stage for models requiring it (NMR2Struct, NMIRacle, and Spec2Mol). Extended results for this stage are provided in Appendix~\ref{app:pre-training}.
Table~\ref{tab:spec2struct} summarizes performance on the spectra-to-molecule generation task across multiple molecular-level metrics, with formal metric definitions given in Appendix~\ref{app:metrics}.
All evaluations follow an enantiomer-aware protocol: predicted and reference molecules are considered equivalent if their canonical SMILES strings match exactly or correspond to enantiomeric (mirror-image) configurations. Details of the evaluation procedure are described in Appendix~\ref{app:eval_criteria}.
\paragraph{Fragments-to-molecule}
Table \ref{tab:sub2struct} reports fragment-to-molecule reconstruction performance on $10,000$ molecules sampled from the pre-training test set. We compare three distinct pre-training strategies: (i) fragment-based reconstruction with binary fragment indicators (NMR2Struct), (ii) fragment-based reconstruction with count-aware fragment representations (ours), and (iii) SMILES-to-SMILES reconstruction using an autoencoder (Spec2Mol). Fragment-based pre-training poses a fundamentally different and more undetermined reconstruction problem than SMILES autoencoding. Given a set of fragments, the model must infer atom-level connectivity and global molecular topology, none of which are explicitly specified in the input. In terms of fragment-based reconstruction paradigm, incorporating fragment occurrences consistently improves reconstruction performance over binary fragment indicators. Top-1 accuracy increases from 0.63 to 0.70, and Top-10 accuracy from 0.76 to 0.81, while maintaining near-perfect chemical validity.
\begin{wraptable}{r}{0.63\columnwidth}
\caption{Pre-training results on 10,000 test molecules. Results are reported in terms of chemical validity, graph edit distance (MCES), and Top-$k$ accuracies. For continuous metrics, values are reported as mean $\pm$ standard deviation computed across test molecules. * indicates our implementations of baseline approaches.}
\smallskip
\label{tab:sub2struct}
\centering
\scriptsize
{\setlength{\tabcolsep}{5pt} 
\begin{tabular}{lcccccc}
\toprule
\textbf{Pre-training} & \textbf{Valid ($\uparrow$)} & \textbf{MCES ($\downarrow$)} &
\multicolumn{3}{c}{\textbf{Top-$k$ Acc.} ($\uparrow$)}\\
\cmidrule(lr){4-6}
& & & 1 & 10 & 15 \\
\midrule
NMR2Struct* &
\cellcolor{gray!17}$\bm{1.00}$ &
$0.92_{\pm 2.32}$ &
$0.63$ & $0.76$ & $0.81$ \\
NMIRacle (Ours) &
$0.97$ &
\cellcolor{gray!17}$\bm{0.57}_{\pm 2.32}$ &
\cellcolor{gray!17}$\bm{0.70}$ &
\cellcolor{gray!17}$\bm{0.81}$ &
\cellcolor{gray!17}$\bm{0.86}$ \\
\midrule
Spec2Mol* & $1.00$ & $0.07_{\pm 0.67}$ & $0.96$ & $0.98$ & $0.98$ \\
\bottomrule
\end{tabular}%
}
\end{wraptable}
These gains indicate that count-aware fragment representations provide additional, quantitative constraints that help resolve ambiguities in molecular assembly, leading to more faithful reconstructions within the fragment-based paradigm.
In contrast, SMILES autoencoding provides a complete description of the molecular graph, making reconstruction substantially easier. Therefore, as expected, Spec2Mol achieves the highest reconstruction accuracy in this stage.
\begin{table*}[!t]
\caption{
Performance comparison for the spectra-to-molecule task across different spectral combinations.
Results are reported in terms of chemical validity, structural similarity (Tanimoto), graph edit distance (MCES), string-level distance (Levenshtein), and Top-$k$ accuracies. For continuous metrics, values are reported as mean $\pm$ standard deviation computed across test molecules. * indicates our implementations of existing baseline approaches.
}
\label{tab:spec2struct}
\begin{center}
\begin{small}
\begin{sc}
\resizebox{\textwidth}{!}{%
\begin{tabular}{l l c c c c c c c c c c}
\toprule
\multirow{2}{*}{\textbf{Spectra}} &
\multirow{2}{*}{\textbf{Model}} &
\multirow{2}{*}{\textbf{Valid} ($\uparrow$)} &
\multicolumn{3}{c}{\textbf{Tanimoto} ($\uparrow$)} &
\multirow{2}{*}{\textbf{MCES} ($\downarrow$)} &
\multirow{2}{*}{\textbf{Lev.} ($\downarrow$)} &
\multicolumn{4}{c}{\textbf{Top-$k$ Acc.} ($\uparrow$)} \\
\cmidrule(lr){4-6} \cmidrule(l){9-12}
& & & Morgan & MACCS & RDKit & & & 1 & 5 & 10 & 15 \\
\midrule
\multirow[c]{5}{*}{\shortstack{${}^1$H \\ \\ $+ {}^{13}$C-NMR}}
& SMILES Transformer   & $1.00$ & $0.72_{\pm 0.25}$ & $0.90_{\pm 0.13}$ & $0.77_{\pm 0.24}$ & $4.21_{\pm 4.40}$ & $7.01_{\pm 8.33}$ & $0.25$ & $0.31$ & $0.36$ & $0.38$ \\
& SELFIES Transformer  & $1.00$ & $0.61_{\pm 0.26}$ & $0.85_{\pm 0.14}$ & $0.67_{\pm 0.24}$ & $5.70_{\pm 4.57}$ & $10.27_{\pm 9.65}$ & $0.15$ & $0.20$ & $0.23$ & $0.24$ \\
& NMR2Struct*  & $1.00$ & $0.79_{\pm 0.24}$ & $0.93_{\pm 0.11}$ & $0.82_{\pm 0.23}$ & $3.17_{\pm 4.15}$ & $5.26_{\pm 7.58}$ & $0.35$ & $0.41$ & $0.47$ & $0.50$ \\
& Spec2Mol* & $1.00$ & $0.32_{\pm 0.13}$ & $0.69_{\pm 0.12}$ & $0.45_{\pm 0.14}$ & $10.78_{\pm 3.78}$ & $17.76_{\pm 9.61}$ & $0.00$ & $0.00$ & $0.01$ & $0.01$ \\
\cmidrule(lr){2-12}
& NMIRacle (Ours) & $1.00$ &
\cellcolor{gray!17}$\bm{0.82}_{\pm 0.23}$ &
\cellcolor{gray!17}$\bm{0.94}_{\pm 0.11}$ &
\cellcolor{gray!17}$\bm{0.85}_{\pm 0.22}$ &
\cellcolor{gray!17}$\bm{2.72}_{\pm 3.96}$ &
\cellcolor{gray!17}$\bm{4.49}_{\pm 7.13}$ &
\cellcolor{gray!17}$\bm{0.39}$ &
\cellcolor{gray!17}$\bm{0.45}$ &
\cellcolor{gray!17}$\bm{0.52}$ &
\cellcolor{gray!17}$\bm{0.56}$ \\
\midrule

\multirow[c]{5}{*}{\shortstack{IR \\ \\ $+$ ${}^1$H-NMR}}
& SMILES Transformer   & $1.00$ & 0.$77_{\pm 0.24}$ & $0.93_{\pm 0.10}$ & $0.81_{\pm 0.23}$ & $3.45_{\pm 4.11}$ & $5.79_{\pm 7.77}$ & $0.30$ & $0.36$ & $0.42$ & $0.45$ \\
& SELFIES Transformer  & $1.00$ & $0.64_{\pm 0.26}$ & $0.88_{\pm 0.13}$ & $0.70_{\pm 0.13}$ & $5.16_{\pm 4.46}$ & $9.52_{\pm 9.57}$ & $0.18$ & $0.22$ & $0.27$ & $0.28$ \\
& NMR2Struct* & $1.00$ & $0.82_{\pm 0.23}$ & $0.95_{\pm 0.09}$ & $0.85_{\pm 0.21}$ & $2.71_{\pm 3.86}$ & $4.09_{\pm 6.65}$ & $0.38$ & $0.44$ & $0.51$ & $0.54$ \\
& Spec2Mol* & $1.00$ & $0.29_{\pm 0.12}$ & $0.67_{\pm 0.12}$ & $0.42_{\pm 0.13}$ & $10.70_{\pm 3.78}$ & $17.77_{\pm 8.91}$ & $0.00$ & $0.00$ & $0.00$ & $0.00$ \\
\cmidrule(lr){2-12}
& NMIRacle (Ours) & $1.00$ &
\cellcolor{gray!17}$\bm{0.86}_{\pm 0.21}$ &
\cellcolor{gray!17}$\bm{0.96}_{\pm 0.08}$ &
\cellcolor{gray!17}$\bm{0.89}_{\pm 0.19}$ &
\cellcolor{gray!17}$\bm{2.06}_{\pm 3.49}$ &
\cellcolor{gray!17}$\bm{3.52}_{\pm 6.50}$ &
\cellcolor{gray!17} $\bm{0.45}$ &
\cellcolor{gray!17}$\bm{0.50}$ &
\cellcolor{gray!17}$\bm{0.59}$ &
\cellcolor{gray!17}$\bm{0.63}$ \\
\midrule

\multirow[c]{5}{*}{\shortstack{IR \\[0.3em] $+ {}^1$H-NMR \\ \\ $+ {}^{13}$C-NMR}}
& SMILES Transformer   & $1.00$ & $0.76_{\pm 0.22}$ & $0.93_{\pm 0.10}$ & $0.80_{\pm 0.22}$ & $3.59_{\pm 4.09}$ & $6.22_{\pm 8.04}$ & $0.28$ & $0.34$ & $0.40$ & $0.42$ \\
& SELFIES Transformer  & $1.00$ & $0.64_{\pm 0.26}$ & $0.88_{\pm 0.12}$ & $0.71_{\pm 0.24}$ & $5.07_{\pm 4.39}$ & $9.48_{\pm 9.50}$ & $0.17$ & $0.22$ & $0.26$ & $0.28$ \\
& NMR2Struct* & $1.00$ & $0.84_{\pm 0.22}$ & $0.96_{\pm 0.09}$ & $0.87_{\pm 0.20}$ & $2.39_{\pm 3.67}$ & $4.18_{\pm 6.97}$ & $0.41$ & $0.47$ & $0.55$ & $0.58$ \\
& Spec2Mol* & $1.00$ & $0.35_{\pm 0.15}$ & $0.72_{\pm 0.12}$ & $0.47_{\pm 0.15}$ & $9.85_{\pm 4.28}$ & $17.45_{\pm 9.94}$ & $0.00$ & $0.01$ & $0.01$ & $0.01$ \\
\cmidrule(lr){2-12}
& NMIRacle (Ours) & $1.00$ &
\cellcolor{gray!17}$\bm{0.88}_{\pm 0.20}$ &
\cellcolor{gray!17}$\bm{0.97}_{\pm 0.07}$ &
\cellcolor{gray!17}$\bm{0.90}_{\pm 0.18}$ &
\cellcolor{gray!17}$\bm{1.82}_{\pm 3.29}$ &
\cellcolor{gray!17}$\bm{3.21}_{\pm 6.23}$ &
\cellcolor{gray!17}$\bm{0.48}$ &
\cellcolor{gray!17}$\bm{0.53}$ &
\cellcolor{gray!17}$\bm{0.61}$ &
\cellcolor{gray!17}$\bm{0.66}$ \\
\midrule

\multirow[c]{5}{*}{\shortstack{IR \\ \\ $+ {}^{13}$C-NMR}}
& SMILES Transformer   & $1.00$ & $0.55_{\pm 0.25}$ & $0.85_{\pm 0.13}$ & $0.64_{\pm 0.22}$ & $6.37_{\pm 4.31}$ & $11.94_{\pm 9.83}$ & $0.09$ & $0.12$ & $0.14$ & $0.15$ \\
& SELFIES Transformer  & $1.00$ & $0.47_{\pm 0.23}$ & $0.81_{\pm 0.13}$ & $0.58_{\pm 0.21}$ & $7.22_{\pm 4.22}$ & $13.83_{\pm 10.10}$ & $0.05$ & $0.08$ & $0.09$ & $0.10$ \\
& NMR2Struct* & $1.00$ & $0.59_{\pm 0.26}$ & $0.87_{\pm 0.13}$ & $0.68_{\pm 0.23}$ & $5.71_{\pm 4.39}$ & $10.84_{\pm 9.84}$ & $0.12$ & $0.16$ & $0.20$ & $0.21$ \\
& Spec2Mol* & $1.00$ & $0.35_{\pm 0.15}$ & $0.73_{\pm 0.12}$ & $0.47_{\pm 0.15}$ & $9.42_{\pm 4.19}$ & $17.05_{\pm 10.14}$ & $0.00$ & $0.01$ & $0.01$ & $0.01$ \\
\cmidrule(lr){2-12}
& NMIRacle (Ours) & $1.00$ &
\cellcolor{gray!17}$\bm{0.63}_{\pm 0.26}$ &
\cellcolor{gray!17}$\bm{0.88}_{\pm 0.12}$ &
\cellcolor{gray!17}$\bm{0.71}_{\pm 0.23}$ &
\cellcolor{gray!17}$\bm{5.22}_{\pm 4.38}$ &
\cellcolor{gray!17}$\bm{9.84}_{\pm 9.62}$ &
\cellcolor{gray!17} $\bm{0.14}$ &
\cellcolor{gray!17}$\bm{0.19}$ &
\cellcolor{gray!17}$\bm{0.23}$ &
\cellcolor{gray!17}$\bm{0.24}$ \\
\bottomrule
\end{tabular}
}
\end{sc}
\end{small}
\end{center}
\end{table*}
\begin{figure*}[!t]
\centering
\includegraphics[width=0.99\textwidth]{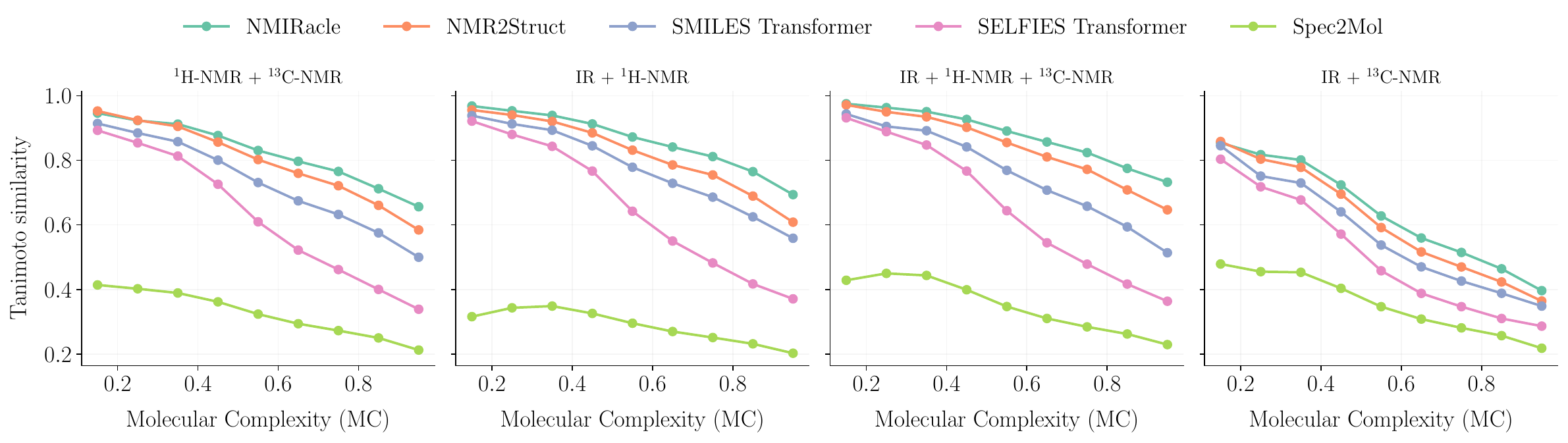}
\caption{
Model performance across molecular complexity bins for different spectral combinations.
NMIRacle maintains higher Tanimoto similarity even for structurally-rich molecules.
}
\label{fig:complexity}
\end{figure*}
\paragraph{Spectra-to-molecule}
Results for molecular generation from multi-spectra inputs are reported in Table \ref{tab:spec2struct}. Across all evaluated spectral combinations, NMIRacle consistently achieves the strongest performance. When all three modalities (IR, ${}^{1}$H-NMR, ${}^{13}$C-NMR) are available, NMIRacle attains a Top-1 accuracy of $0.48$ and a Top-15 accuracy of $0.66$, outperforming NMR2Struct ($0.41$ / $0.58$) and all other baselines.
Models that rely on more constrained or less aligned intermediate representations perform substantially worse. For instance, SELFIES-based transformer exhibits weaker performance: while SELFIES guarantees chemical validity, it may also reduce flexibility in conditional generative modeling, leading to worse performance on other structural metrics~\citep{skinnider2024invalid}, and making it harder for the model to resolve structural ambiguities from IR and NMR spectra. Spec2Mol also underperforms in the spectra-to-molecule task, despite its strong pre-training results. This highlights a limitation of SMILES autoencoding as a pre-training strategy for multi-spectra conditional generation: in our setting, aligning spectral representations with a continuous SMILES latent space proves significantly more difficult than conditioning generation on an intermediate, chemically-grounded fragment representation. Similar trends have been observed in recent spectra-to-molecule benchmarks~\citep{bohde2025diffms}. Additional qualitative success cases are reported in Appendix~\ref{app:success_cases}, illustrating molecules that are correctly recovered by NMIRacle while all competing methods fail. 
\paragraph{Scaling to more complex molecules}
To assess models' robustness with respect to molecular size and structural diversity, we introduce
an empirical \emph{molecular complexity} (MC) index:

\begin{equation}
MC := \frac{1}{3}\left(\frac{N_{h}}{N_{h_{\max}}} + \frac{N_{u}}{N_{u_{\max}}} + \frac{N_{r}}{N_{r_{\max}}}\right),
\label{eq:mc}
\end{equation}

where $N_h$, $N_u$, and $N_r$ denote the number of heavy atoms, unique elements, and rings, respectively. $N_{(\cdot)_{\max}}$ corresponds to the $99$-th percentile of the corresponding distribution.
Molecules are partitioned into ten complexity bins, and Tanimoto similarity (based on Morgan fingerprints) is reported per bin in Figure~\ref{fig:complexity} across different spectral combinations.
NMIRacle consistently outperforms baseline models across all complexity bins, maintaining a stable performance lead even as the overall Tanimoto similarity declines for structurally-rich molecules.
\subsection{Summary of additional studies}
We provide in Appendix~\ref{app} additional analyses that further characterize the proposed framework. In particular:
\begin{itemize}
    \item In Appendix~\ref{app:pre-training} we report extended results for the pre-training stage, including additional molecular similarity metrics.

    \item In Appendix~\ref{app:ablations} we present ablation studies isolating the contribution of learnable positional encodings and explicit inter-modal attention, showing that both components yield consistent performance gains.

    \item In Appendix~\ref{app:eval_criteria} we detail the enantiomer-aware evaluation protocol adopted in the main results. We further examine alternative evaluation criteria that relax stereochemical constraints, showing that a non-negligible fraction of errors arises from stereochemical ambiguity rather than incorrect connectivity.
    
    \item In Appendix~\ref{app:fail_modes} we provide a detailed analysis of model failure cases, identifying fragment misprediction as the dominant source of error across different spectral settings.

    \item In Appendix~\ref{app:frag_pred} we analyze fragment occurrence prediction under the multi-task objective (Eq.~\ref{eq:stage2_loss}), showing that accurately predicted fragments rapidly cover most of the fragment occurrence space, highlighting practical utility for common structural motifs.
    
\end{itemize}

\section{Conclusion}
Motivated by recent advances in spectra-to-molecule machine learning, we present NMIRacle, a two-stage generative framework for molecular structure elucidation from combinations of raw IR, $^{1}$H-NMR, and ${}^{13}$C-NMR spectra. Our framework builds upon previous efforts towards spectra-to-molecule modeling with minimal assumptions.
Our approach combines a count-aware fragment prior with a hierarchical, multi-spectra encoder, enabling informative spectral conditioning. Across multiple evaluation settings, NMIRacle achieves consistently strong molecular elucidation performance and exhibits robust generalization to structurally-complex molecules.
Overall, NMIRacle provides a flexible foundation for realistic, data-driven molecular elucidation from spectral evidence. 
Additional discussion related to limitations and future directions is provided in~\ref{app:outlook}.

\section{Acknowledgments}
The authors acknowledge the AI for Chemistry: AIchemy
hub for funding (EPSRC grant EP/Y028775/1 and
EP/Y028759/1).

\bibliography{iclr2026_conference}

@article {kretschmer2023mces,
	author = {Kretschmer, Fleming and Seipp, Jan and Ludwig, Marcus and Klau, Gunnar W. and B{\"o}cker, Sebastian},
	title = {Small molecule machine learning: All models are wrong, some may not even be useful},
	elocation-id = {2023.03.27.534311},
	year = {2023},
	doi = {10.1101/2023.03.27.534311},
	publisher = {Cold Spring Harbor Laboratory},
	journal = {bioRxiv}
}

@inproceedings{alberts2023learning,
  title={Learning the language of \text{NMR}: structure elucidation from \text{NMR} spectra using transformer models},
  author={Alberts, Marvin and Zipoli, Federico and Vaucher, Alain},
  booktitle={AI for Accelerated Materials Design-NeurIPS 2023 Workshop},
  year={2023}
}

@article{yang2026diffnmr,
  title={Diff{NMR}: diffusion models for nuclear magnetic resonance spectra elucidation},
  author={Yang, Qingsong and Wu, Binglan and Liu, Xuwei and Chen, Bo and Li, Wei and Long, Gen and Chen, Xin and Xiao, Mingjun},
  journal={Materials Futures},
  volume={5},
  number={1},
  pages={015601},
  year={2026},
  publisher={IOP Publishing}
}

@article{jin2025nmrsolver,
  title={{NMR-Solver}: Automated Structure Elucidation via Large-Scale Spectral Matching and Physics-Guided Fragment Optimization},
  author={Jin, Yongqi and Wang, Jun-Jie and Xu, Fanjie and Ji, Xiaohong and Gao, Zhifeng and Zhang, Linfeng and Ke, Guolin and Zhu, Rong and others},
  journal={arXiv preprint arXiv:2509.00640},
  year={2025}
}

@article{wang2025diffspectra,
  title={Diff{S}pectra: Molecular structure elucidation from spectra using diffusion models},
  author={Wang, Liang and Rong, Yu and Xu, Tingyang and Zhong, Zhenyi and Liu, Zhiyuan and Wang, Pengju and Zhao, Deli and Liu, Qiang and Wu, Shu and Zhang, Yang},
  journal={arXiv preprint arXiv:2507.06853},
  year={2025}
}

@article{mestrelab_mnova,
author = {Willcott, Mark Robert},
doi = {10.1021/ja906709t},
isbn = {0002-7863},
journal = {Journal of the American Chemical Society},
number = {36},
pages = {13180--13180},
publisher = {American Chemical Society},
title = {Mest{R}e Nova},
type = {doi: 10.1021/ja906709t},
url = {https://doi.org/10.1021/ja906709t},
volume = {131},
year = {2009},
}

@book{clayden2012organic,
  title     = {Organic Chemistry},
  author    = {Clayden, Jonathan and Greeves, Nick and Warren, Stuart and Wothers, Peter},
  edition   = {2},
  chapter   = {13},
  year      = {2012},
  publisher = {Oxford University Press},
}

@misc{rdkit,
  title  = {{RDKit}: Open-source cheminformatics},
  author = {Landrum, Gregory},
  url    = {https://www.rdkit.org},
}

@article{skinnider2024invalid,
	author = {Skinnider, Michael A. },
	date = {2024/04/01},
	doi = {10.1038/s42256-024-00821-x},
	id = {Skinnider2024},
	isbn = {2522-5839},
	journal = {Nature Machine Intelligence},
	number = {4},
	pages = {437--448},
	title = {Invalid SMILES are beneficial rather than detrimental to chemical language models},
	url = {https://doi.org/10.1038/s42256-024-00821-x},
	volume = {6},
	year = {2024}}

@article{mirza2024elucidating,
	title        = {Elucidating Structures from Spectra Using Multimodal Embeddings and Discrete Optimization},
	author       = {Mirza, Adrian and Jablonka, Kevin Maik},
	year         = {2024},
	journal      = {ChemRxiv},
	doi          = {10.26434/chemrxiv-2024-f3b18-v2},
}

@inproceedings{fang2024domain,
title={Domain-Agnostic Molecular Generation with Chemical Feedback},
author={Yin Fang and Ningyu Zhang and Zhuo Chen and Lingbing Guo and Xiaohui Fan and Huajun Chen},
booktitle={The Twelfth International Conference on Learning Representations},
year={2024},
url={https://openreview.net/forum?id=9rPyHyjfwP}
}

@inproceedings{wang2025madgen,
title={{MADGEN}: Mass-Spec attends to De Novo Molecular generation},
author={Yinkai Wang and Xiaohui Chen and Liping Liu and Soha Hassoun},
booktitle={The Thirteenth International Conference on Learning Representations},
year={2025},
url={https://openreview.net/forum?id=78tc3EiUrN}
}

@article{hu2024accurate,
	author = {Hu, Frank and Chen, Michael S. and Rotskoff, Grant M. and Kanan, Matthew W. and Markland, Thomas E.},
	date = {2024/11/27},
	doi = {10.1021/acscentsci.4c01132},
	isbn = {2374-7943},
	journal = {ACS Central Science},
	journal1 = {ACS Central Science},
	journal2 = {ACS Cent. Sci.},
	month = {11},
	number = {11},
	pages = {2162--2170},
	publisher = {American Chemical Society},
	title = {Accurate and Efficient Structure Elucidation from Routine One-Dimensional {NMR} Spectra Using Multitask Machine Learning},
	type = {doi: 10.1021/acscentsci.4c01132},
	url = {https://doi.org/10.1021/acscentsci.4c01132},
	volume = {10},
	year = {2024},
	year1 = {2024}}

@inproceedings{zhu20243m,
title={{3M}-{D}iffusion: Latent Multi-Modal Diffusion for Language-Guided Molecular Structure Generation},
author={Huaisheng Zhu and Teng Xiao and Vasant G Honavar},
booktitle={First Conference on Language Modeling},
year={2024},
url={https://openreview.net/forum?id=DomBynQsqt}
}

@inproceedings{xie2021mars,
title={{MARS}: Markov Molecular Sampling for Multi-objective Drug Discovery},
author={Yutong Xie and Chence Shi and Hao Zhou and Yuwei Yang and Weinan Zhang and Yong Yu and Lei Li},
booktitle={International Conference on Learning Representations},
year={2021},
url={https://openreview.net/forum?id=kHSu4ebxFXY}
}

@inproceedings{kim2025molllama,
title={Mol-{LL}a{MA}: Towards General Understanding of Molecules in Large Molecular Language Model},
author={Dongki Kim and Wonbin Lee and Sung Ju Hwang},
booktitle={NeurIPS 2025 AI for Science Workshop},
year={2025},
url={https://openreview.net/forum?id=TTeQ3bOwsL}
}

@article{lee2025fragfm,
  title={Frag{FM}: Efficient Fragment-Based Molecular Generation via Discrete Flow Matching},
  author={Lee, Joongwon and Kim, Seonghwan and Kim, Wou Youn},
  journal={arXiv preprint arXiv:2502.15805},
  year={2025}
}

@article{liu2024gitmol,
   title={GIT-Mol: A multi-modal large language model for molecular science with graph, image, and text},
   volume={171},
   ISSN={0010-4825},
   url={http://dx.doi.org/10.1016/j.compbiomed.2024.108073},
   DOI={10.1016/j.compbiomed.2024.108073},
   journal={Computers in Biology and Medicine},
   publisher={Elsevier BV},
   author={Liu, Pengfei and Ren, Yiming and Tao, Jun and Ren, Zhixiang},
   year={2024},
   month=mar, pages={108073} }

@InProceedings{jin2020hierarchical,
  title = 	 {Hierarchical Generation of Molecular Graphs using Structural Motifs},
  author =       {Jin, Wengong and Barzilay, Dr.Regina and Jaakkola, Tommi},
  booktitle = 	 {Proceedings of the 37th International Conference on Machine Learning},
  pages = 	 {4839--4848},
  year = 	 {2020},
  editor = 	 {III, Hal Daumé and Singh, Aarti},
  volume = 	 {119},
  series = 	 {Proceedings of Machine Learning Research},
  publisher =    {PMLR},
  url = 	 {https://proceedings.mlr.press/v119/jin20a.html}
}

@article{tom2025stereo,
    author = {Tom, Gary and Yu, Edwin and Yoshikawa, Naruki and Jorner, Kjell and Aspuru-Guzik, Alán},
    title = {Stereochemistry-aware string-based molecular generation},
    journal = {PNAS Nexus},
    volume = {4},
    number = {11},
    pages = {pgaf329},
    year = {2025},
    issn = {2752-6542},
    doi = {10.1093/pnasnexus/pgaf329},
    url = {https://doi.org/10.1093/pnasnexus/pgaf329},
    eprint = {https://academic.oup.com/pnasnexus/article-pdf/4/11/pgaf329/64703722/pgaf329.pdf},
}

@InProceedings{jin2018junction,
  title = 	 {Junction Tree Variational Autoencoder for Molecular Graph Generation},
  author = {Jin, Wengong and Barzilay, Regina and Jaakkola, Tommi},
  booktitle = 	 {Proceedings of the 35th International Conference on Machine Learning},
  pages = 	 {2323--2332},
  year = 	 {2018},
  editor = 	 {Dy, Jennifer and Krause, Andreas},
  volume = 	 {80},
  series = 	 {Proceedings of Machine Learning Research},
  month = 	 {10--15 Jul},
  publisher =    {PMLR},
  url = 	 {https://proceedings.mlr.press/v80/jin18a.html}
}

@inproceedings{liu2024graphdit,
 author = {Liu, Gang and Xu, Jiaxin and Luo, Tengfei and Jiang, Meng},
 booktitle = {Advances in Neural Information Processing Systems},
 doi = {10.52202/079017-0260},
 editor = {A. Globerson and L. Mackey and D. Belgrave and A. Fan and U. Paquet and J. Tomczak and C. Zhang},
 pages = {8065--8092},
 publisher = {Curran Associates, Inc.},
 title = {Graph Diffusion Transformers for Multi-Conditional Molecular Generation},
 volume = {37},
 year = {2024}
}

@inproceedings{pei20253dmolt5,
title={{3D-MolT5}: Leveraging Discrete Structural Information for Molecule-Text Modeling},
author={Qizhi Pei and Rui Yan and Kaiyuan Gao and Jinhua Zhu and Lijun Wu},
booktitle={The Thirteenth International Conference on Learning Representations},
year={2025},
url={https://openreview.net/forum?id=eGqQyTAbXC}
}

@misc{spectrabase,
  author = {{John Wiley \& Sons, Inc.}},
  title = {{SpectraBase}},
  howpublished = {\url{https://spectrabase.com}},
}

@inproceedings{christofidellis2023unifying,
  title={Unifying molecular and textual representations via multi-task language modelling},
  author={Christofidellis, Dimitrios and Giannone, Giorgio and Born, Jannis and Winther, Ole and Laino, Teodoro and Manica, Matteo},
  booktitle={International Conference on Machine Learning},
  pages={6140--6157},
  year={2023},
  organization={PMLR}
}

@inproceedings{bohde2025diffms,
title={Diff{MS}: Diffusion Generation of Molecules Conditioned on Mass Spectra},
author={Montgomery Bohde and Mrunali Manjrekar and Runzhong Wang and Shuiwang Ji and Connor W. Coley},
booktitle={Forty-second International Conference on Machine Learning},
year={2025},
url={https://openreview.net/forum?id=EvILcv2v8L}
}

@article{yang2021crees,
	author = {Yang, Zhuo and Song, Jianfei and Yang, Minjian and Yao, Lin and Zhang, Jiahua and Shi, Hui and Ji, Xiangyang and Deng, Yafeng and Wang, Xiaojian},
	date = {2021/12/21},
	doi = {10.1021/acs.analchem.1c04307},
	isbn = {0003-2700},
	journal = {Analytical Chemistry},
	journal1 = {Analytical Chemistry},
	journal2 = {Anal. Chem.},
	month = {12},
	number = {50},
	pages = {16947--16955},
	publisher = {American Chemical Society},
	title = {Cross-Modal Retrieval between ${}^{13}${C NMR} Spectra and Structures for Compound Identification Using Deep Contrastive Learning},
	type = {doi: 10.1021/acs.analchem.1c04307},
	url = {https://doi.org/10.1021/acs.analchem.1c04307},
	volume = {93},
	year = {2021},
	year1 = {2021}}

@article{litsa2021spec2mol,
	author = {Litsa, Eleni E. and Chenthamarakshan, Vijil and Das, Payel and Kavraki, Lydia E.},
	date = {2023/06/23},
	doi = {10.1038/s42004-023-00932-3},
	id = {Litsa2023},
	isbn = {2399-3669},
	journal = {Communications Chemistry},
	number = {1},
	pages = {132},
	title = {An end-to-end deep learning framework for translating mass spectra to de-novo molecules},
	url = {https://doi.org/10.1038/s42004-023-00932-3},
	volume = {6},
	year = {2023}}

@article{ruddigkeit2012enumeration,
	author = {Ruddigkeit, Lars and van Deursen, Ruud and Blum, Lorenz C. and Reymond, Jean-Louis},
	date = {2012/11/26},
	doi = {10.1021/ci300415d},
	isbn = {1549-9596},
	journal = {Journal of Chemical Information and Modeling},
	journal1 = {Journal of Chemical Information and Modeling},
	journal2 = {J. Chem. Inf. Model.},
	month = {11},
	number = {11},
	pages = {2864--2875},
	publisher = {American Chemical Society},
	title = {Enumeration of 166 Billion Organic Small Molecules in the Chemical Universe Database {GDB-17}},
	type = {doi: 10.1021/ci300415d},
	url = {https://doi.org/10.1021/ci300415d},
	volume = {52},
	year = {2012},
	year1 = {2012}}

@article{polishchuk2013molsize,
	author = {Polishchuk, P. G. and Madzhidov, T. I. and Varnek, A.},
	date = {2013/08/01},
	doi = {10.1007/s10822-013-9672-4},
	id = {Polishchuk2013},
	isbn = {1573-4951},
	journal = {Journal of Computer-Aided Molecular Design},
	number = {8},
	pages = {675--679},
	title = {Estimation of the size of drug-like chemical space based on {GDB-17} data},
	url = {https://doi.org/10.1007/s10822-013-9672-4},
	volume = {27},
	year = {2013}}

@article{yao2023cmgnet,
	author = {Yao, Lin and Yang, Minjian and Song, Jianfei and Yang, Zhuo and Sun, Hanyu and Shi, Hui and Liu, Xue and Ji, Xiangyang and Deng, Yafeng and Wang, Xiaojian},
	date = {2023/03/28},
	doi = {10.1021/acs.analchem.2c05817},
	isbn = {0003-2700},
	journal = {Analytical Chemistry},
	journal1 = {Analytical Chemistry},
	journal2 = {Anal. Chem.},
	month = {03},
	number = {12},
	pages = {5393--5401},
	publisher = {American Chemical Society},
	title = {Conditional Molecular Generation Net Enables Automated Structure Elucidation Based on ${}^{13}${C NMR} Spectra and Prior Knowledge},
	type = {doi: 10.1021/acs.analchem.2c05817},
	url = {https://doi.org/10.1021/acs.analchem.2c05817},
	volume = {95},
	year = {2023},
	year1 = {2023}}

@article{guo2024can,
	title={Can \text{LLMs} solve molecule puzzles? a multimodal benchmark for molecular structure elucidation},
	author={Guo, Kehan and Nan, Bozhao and Zhou, Yujun and Guo, Taicheng and Guo, Zhichun and Surve, Mihir and Liang, Zhenwen and Chawla, Nitesh and Wiest, Olaf and Zhang, Xiangliang},
	journal={Advances in Neural Information Processing Systems},
	volume={37},
	pages={134721--134746},
	year={2024}
}

@article{bushuiev2024massspecgym,
	title={MassSpecGym: A benchmark for the discovery and identification of molecules},
	author={Bushuiev, Roman and Bushuiev, Anton and de Jonge, Niek and Young, Adamo and Kretschmer, Fleming and Samusevich, Raman and Heirman, Janne and Wang, Fei and Zhang, Luke and D{\"u}hrkop, Kai and others},
	journal={Advances in Neural Information Processing Systems},
	volume={37},
	pages={110010--110027},
	year={2024}
}

@article{alberts2024unraveling,
	title={Unraveling molecular structure: A multimodal spectroscopic dataset for chemistry},
	author={Alberts, Marvin and Schilter, Oliver and Zipoli, Federico and Hartrampf, Nina and Laino, Teodoro},
	journal={Advances in Neural Information Processing Systems},
	volume={37},
	pages={125780--125808},
	year={2024}
}

@article{edwards2022translation,
	title={Translation between molecules and natural language},
	author={Edwards, Carl and Lai, Tuan and Ros, Kevin and Honke, Garrett and Cho, Kyunghyun and Ji, Heng},
	journal={arXiv preprint arXiv:2204.11817},
	year={2022}
}
\bibliographystyle{iclr2026_conference}

\newpage
\appendix

\section{Appendix}\label{app}
\subsection{Related work}\label{app:related_work}
\paragraph{Data-driven molecular elucidation from spectroscopy}
Molecular structure elucidation has recently emerged as a benchmark for multimodal AI, with several works addressing the task under diverse settings.
\citet{guo2024can} introduced MolPuzzle, a zero-shot benchmark framing structure elucidation as a multi-step reasoning task integrating spectral analysis, property inference, and functional groups assembly.
MassSpecGym~\citep{bushuiev2024massspecgym} provides standardized metrics and curated datasets for molecular \textit{de novo} generation and retrieval from mass spectra. 
\citet{alberts2024unraveling} present a large-scale multimodal dataset of $\sim$790k molecules paired with simulated spectroscopic data, providing a unified benchmark for multimodal structure elucidation.
\citet{mirza2024elucidating} align spectral and molecular embeddings via contrastive learning for cross-modal retrieval, further employing a genetic algorithm to introduce novelty among retrieved molecular candidates.
DiffNMR~\citep{yang2026diffnmr} employs a conditional discrete diffusion model to perform de novo molecular structure elucidation from NMR spectra, iteratively refining the molecular graph structure and using a two-stage pre-training for enhanced spectra-molecule alignment.
Spec2Mol~\citep{litsa2021spec2mol} reconstructs molecular SMILES via a gated recurrent unit (GRU)–based autoencoder, then aligns a convolution-based spectral encoder to the molecular latent space, enabling direct reconstruction from mass spectra. DiffMS~\citep{bohde2025diffms} introduces a discrete diffusion framework for molecular graph generation, in which a graph transformer denoises adjacency matrices conditioned on molecular fingerprints predicted from mass spectra.
NMR2Struct~\citep{hu2024accurate} employs an autoregressive multi-task setup in which a fragment-based generative model is reused for joint SMILES generation and fingerprint prediction conditioned on input spectra. DiffSpectra~\citep{wang2025diffspectra} introduces a diffusion-based framework for elucidating 3D molecules from Ultraviolet-visible (UV-Vis), IR, and Raman spectra, employing an SE(3)-equivariant architecture to jointly infer the 2D topology and 3D geometry of the molecule. \citet{}
Despite these advances, most AI-driven molecular elucidation methods remain single-modality, reliant on pre-processed inputs unavailable from experimental data, and evaluated on small molecules, falling short of realistic, multi-spectra elucidation scenarios.

\paragraph{Conditional generative models for molecules}
Conditional molecular generation represents a central paradigm in AI-driven Chemistry, enabling the targeted design under textual, structural or multi-modal constraints.
Text-driven inverse design via large language models~\citep{edwards2022translation, fang2024domain, christofidellis2023unifying, pei20253dmolt5}, graph-based conditional diffusion models~\citep{liu2024graphdit}, and multi-modal molecular pipelines integrating images, text, and graphs~\citep{zhu20243m, kim2025molllama, liu2024gitmol} have significantly broadened the range of conditioning modalities.
However, these methods rarely incorporate experimental observables. Spectroscopic signals provide physically grounded, high-dimensional evidence of molecular structure, but exhibit instrument-dependent noise and distributions that challenge generic multimodal architectures~\citep{guo2024can}. 

\paragraph{Fragment-based molecular generative modeling}
Motif-level modeling represents a flexible inductive bias for molecular generation, operating over chemically-meaningful molecular fragments rather than individual atoms. Pioneered by methods like JT-VAE~\citep{jin2018junction} and HierVAE~\citep{jin2020hierarchical}, which introduced hierarchical generation tree-structured scaffold representations, the field has advanced to methods such as MARS~\citep{xie2021mars}, which uses GNN-guided Markov Chain Monte Carlo for iterative fragment editing toward multi-objective property optimization, and FragFM~\citep{lee2025fragfm}, which employs a coarse-to-fine autoencoder combining fragment-level graph generation with atom-level reconstruction. Despite their promise, fragment-based generative approaches have rarely been explored in conditional settings, particularly when the conditioning signal consists of experimental observables such as spectroscopic measurements.
\newpage
\subsection{Pre-training results}\label{app:pre-training}
\begin{table*}[!h]
\caption{
Full performance comparison for the pre-training stage across different models.
Results are reported in terms of structural similarity (Tanimoto), graph edit distance (MCES), string-level distance (Levenshtein), and Top-$k$ accuracies. For continuous metrics, values are reported as mean $\pm$ standard deviation computed across test molecules. * indicates our implementations of baseline approaches.
}
\label{tab:sub2struct_full}
\begin{center}
\begin{small}
\begin{sc}
\resizebox{\textwidth}{!}{%
\begin{tabular}{l c c c c c c c c c c}
\toprule
\multirow{2}{*}{\textbf{Model}} &
\multirow{2}{*}{\textbf{Valid} ($\uparrow$)} &
\multicolumn{3}{c}{\textbf{Tanimoto} ($\uparrow$)} &
\multirow{2}{*}{\textbf{MCES} ($\downarrow$)} &
\multirow{2}{*}{\textbf{Lev.} ($\downarrow$)} &
\multicolumn{4}{c}{\textbf{Top-$k$ Acc.} ($\uparrow$)} \\
\cmidrule(lr){3-5} \cmidrule(lr){8-11}
& & Morgan & MACCS & RDKit & & & 1 & 5 & 10 & 15 \\
\midrule

NMR2Struct* & \cellcolor{gray!17}$\bm{1.00}$ & $0.93_{\pm 0.15}$ & \cellcolor{gray!17}$\bm{0.99}_{\pm 0.03}$ & $0.95_{\pm 0.12}$ & $0.92_{\pm 2.32}$ & $1.61_{\pm 4.34}$ & $0.63$ & $0.68$ & $0.76$ & $0.81$ \\
NMIRacle (Ours) & $0.97$ &
\cellcolor{gray!17}$\bm{0.96}_{\pm 0.12}$ &
\cellcolor{gray!17}$\bm{0.99}_{\pm 0.04}$ &
\cellcolor{gray!17}$\bm{0.97}_{\pm 0.10}$ &
\cellcolor{gray!17}$\bm{0.57}_{\pm 2.32}$ &
\cellcolor{gray!17}$\bm{1.07}_{\pm 4.41}$ &
\cellcolor{gray!17}$\bm{0.70}$ &
\cellcolor{gray!17}$\bm{0.73}$ &
\cellcolor{gray!17}$\bm{0.81}$ &
\cellcolor{gray!17}$\bm{0.86}$ \\
\cmidrule(lr){1-11}
Spec2Mol* & $1.00$ & $0.99_{\pm 0.05}$ & $1.00_{\pm 0.01}$ & $1.00_{\pm 0.03}$ & $0.07_{\pm 0.67}$ & $0.12_{\pm 1.31}$ & $0.96$ & $0.97$ & $0.98$ & $0.98$ \\
\bottomrule
\end{tabular}
}
\end{sc}
\end{small}
\end{center}
\end{table*}
\subsection{Ablation studies}\label{app:ablations}
\begin{figure}[!h]
    \centering
    \includegraphics[width=0.98\columnwidth]{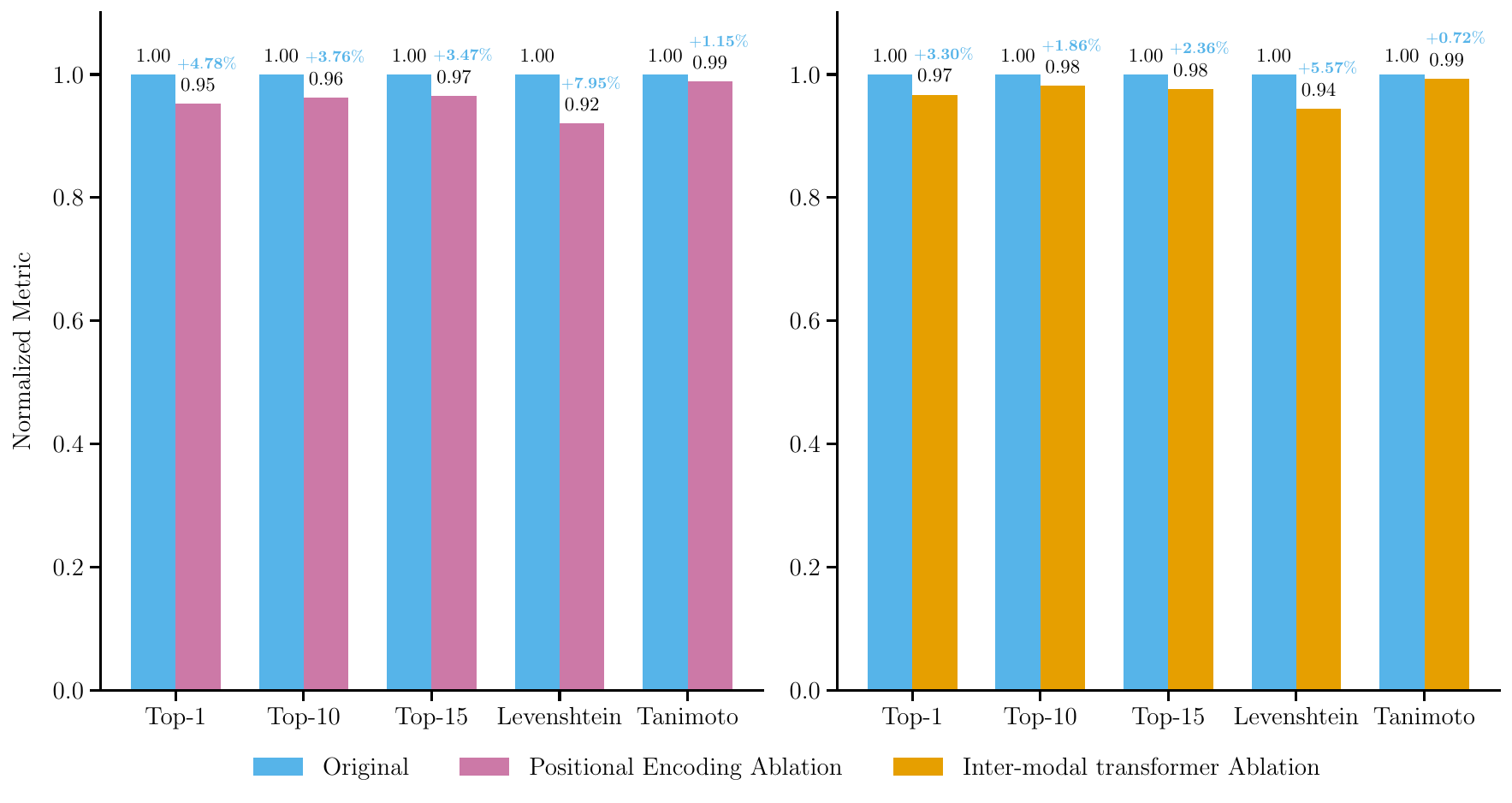}
    \caption{
        Ablation results comparing the impact of learnable positional encodings (left) and inter-modal transformer encoder (\textit{right}).
        Bars report relative performance with respect to the full NMIRacle configuration (blue).
    }
    \label{fig:ablations}
\end{figure}

We conduct ablation experiments to gain further insights on the choice of architectural components. Specifically, we examine: (i) the effect of learnable positional encodings for spectra compared to fixed sinusoidal encodings from prior work~\citep{hu2024accurate}, and (ii) the role of the inter-modal transformer encoder for inter-spectral integration versus a simple concatenation of independently-processed spectra.
As shown in Figure~\ref{fig:ablations}, results are reported as relative performance with respect to the full NMIRacle configuration, which serves as the reference (blue bars). For increasing metrics, relative performance is computed as $\frac{\text{Current value}}{\text{Reference value}}$, and as $\frac{\text{Reference value}}{\text{Current value}}$ for decreasing metrics.
Both components provide consistent improvements, highlighting the benefits of adaptive spectral representation and explicit cross-spectra attention.
\subsection{Evaluation criteria}\label{app:eval_criteria}
\begin{figure}[!t]
    \centering    
    \includegraphics[width=0.98\textwidth]{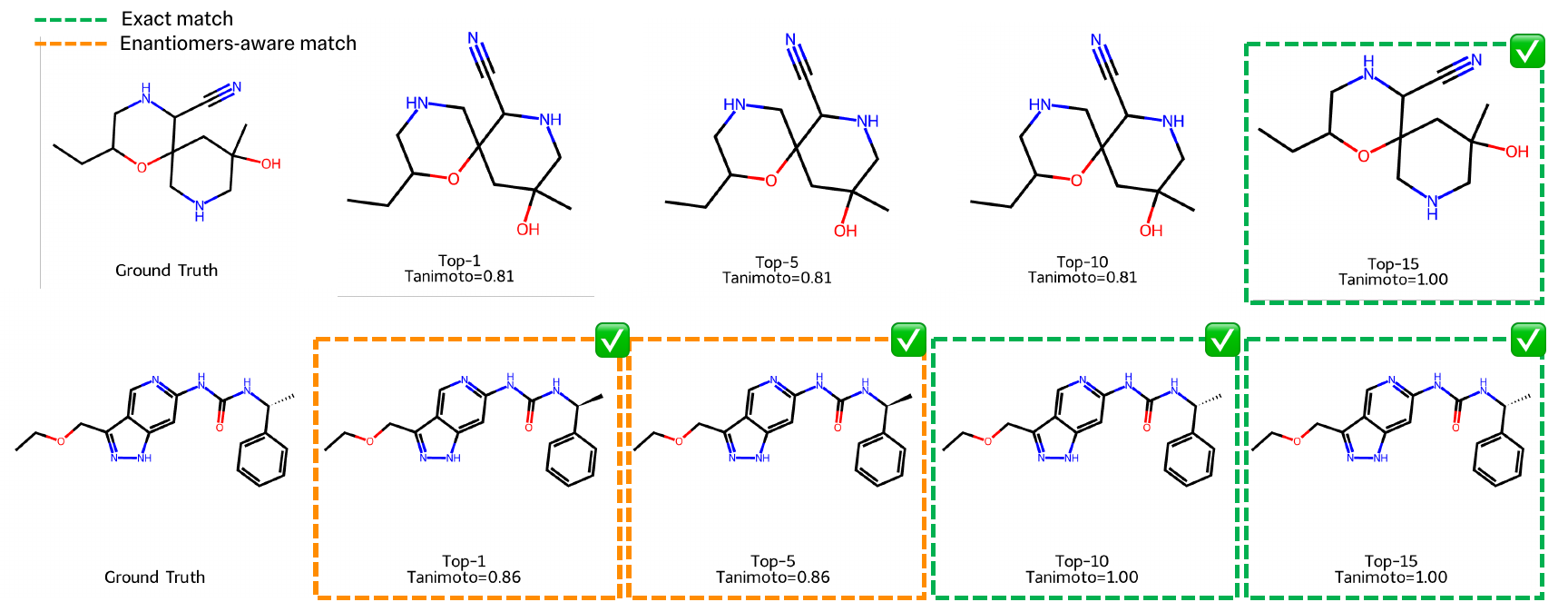}
    \caption{
    Illustration of the enantiomer-aware evaluation scheme. 
    (a) If the generated and reference molecules share identical canonical SMILES, the prediction is counted as an (exact) match (green). 
    (b) If the generated molecule represents the enantiomer of the reference (i.e., a mirror-image configuration), it is likewise treated as an (equivalent) match (orange). 
    This criterion reflects the fact that IR and NMR spectra generally cannot resolve absolute stereochemistry.
    }
    \label{fig:enantiomer_example}
\end{figure}

\begin{figure}[!h]
    \centering
    \includegraphics[width=0.98\textwidth]{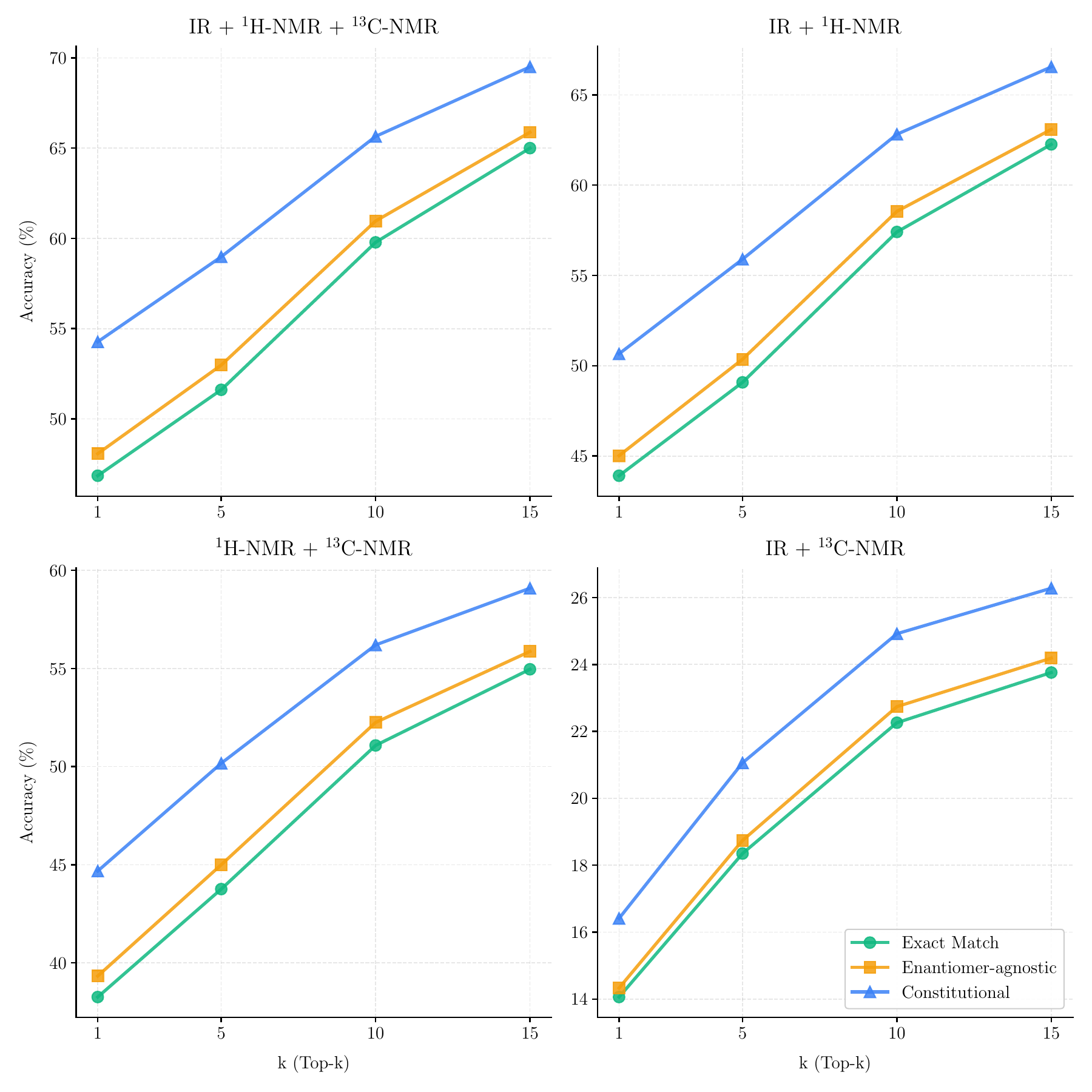}
    \caption{
    Comparison of molecular generation performance under different evaluation criteria, reported as top-$k$ accuracy for various spectral combinations. 
    Results highlight how relaxing stereochemical constraints (from exact, to enantiomer-aware, to constitutional) affects the Top-$k$ for molecular generation.
    }
    \label{fig:eval_protocols}
\end{figure}

In our main results we utilize an enantiomer-aware evaluation protocol. We adopt this scheme because standard IR and NMR spectra are inherently agnostic to absolute stereochemistry, thus making it chemically infeasible to demand full stereochemical resolution. As illustrated in Figure~\ref{fig:enantiomer_example}, this protocol considers a prediction correct if it is either an exact match or the enantiomer (mirror-image configuration) of the ground truth.
To analyze the distribution of mismatches across different levels of structural fidelity, we compare the model's performance under various evaluation criteria, as detailed in Figure~\ref{fig:eval_protocols}.
Three distinct protocols are considered:
(i) \emph{exact match} requires a perfect string-level correspondence between generated and ground-truth SMILES, including atom ordering and stereochemical specification; (ii) \emph{enantiomer-aware} evaluation, adopted as the primary protocol in the main text, relaxes this constraint by treating enantiomeric molecules (i.e., mirror images with identical connectivity) as equivalent, reflecting the limited chirality sensitivity of IR and NMR spectra; (iii) \emph{constitutional} criterion represents the most 'permissive' case, obtained by recomputing ground truth and generated SMILES with \texttt{RDKit}~\citep{rdkit}, using \texttt{Chem.MolToSmiles(isomeric=False)} function.
This operation canonicalizes molecules solely by their bonding topology, disregarding stereochemical information, and thus evaluates whether the predicted structure matches the correct constitutional framework.
We observe consistent improvements under the constitutional metric, highlighting that a fraction of mismatches arise from stereochemical ambiguities rather than errors in molecular connectivity. 
For instance, top-15 accuracy increases from $0.66$ to $0.69$ for the IR + ${}^1$H-NMR + ${}^{13}$C-NMR combination, and from $0.56$ to $0.59$ for ${}^1$H-NMR + ${}^{13}$C-NMR. 
These results suggest potential practical value when the goal is structure elucidation up to constitutional isomerism, rather than full stereochemical resolution.

\subsection{Metrics}\label{app:metrics}
We evaluate model performance through metrics that capture both generation quality and molecular similarity with respect to the corresponding ground truth. For each input spectra $\mathcal{S}$, the model produces a ranked set of $k$ molecular candidates
$\hat{\mathcal{Y}}_k = \{\hat{\mathbf{y}}_1, \dots, \hat{\mathbf{y}}_k\}$
sampled from $p_\theta(\mathbf{y}\mid\mathcal{S})$ and ranked by their average, per-token log-likelihood under the model.
\paragraph{Validity} Fraction of generated molecules that satisfy basic chemical constraints (e.g., valid atom valences). Validity is computed using RDKit’s sanitization routines:
\begin{equation*}
\text{Validity} = \mathbb{E}_{\mathcal{S} \sim \mathcal{D}} \left[ \mathbbm{1}\left\{\exists\, \hat{\mathbf{y}} \in \hat{\mathcal{Y}}_k : \hat{\mathbf{y}} \text{ is chemically valid}\right\}\right].
\end{equation*}
A score of $1$ indicates that at least one valid molecule is generated for every spectra input.

\paragraph{Top-$k$ Accuracy} Measures whether the ground truth SMILES $\mathbf{y}$ appears among the top-$k$ generated candidates for a given set of input spectra $\mathcal{S}$. This captures the model’s ability to exactly recover the target structure when allowed multiple guesses:
\begin{equation*}
\text{Top-}k\,\text{Acc} = 
\mathbb{E}_{(\mathcal{S}, \mathbf{y}) \sim \mathcal{D}}
\left[ \mathbbm{1}\!\left(\exists\, \hat{\mathbf{y}} \in \hat{\mathcal{Y}}_k : \hat{\mathbf{y}} = \mathbf{y}\right) \right].
\end{equation*}

\paragraph{Maximum Common Edge Subgraph (MCES)}
We employ the graph edit distance between a predicted molecule and the corresponding ground truth, following the implementation by \citet{kretschmer2023mces}. Given that we denote this distance by $d_{\text{mces}}$, then the corresponding metric is calculated as:
\begin{equation*}
\text{MCES} = \mathbb{E}_{(\mathcal{S}, \mathbf{y}) \sim \mathcal{D}} \left[ \min_{\hat{\mathbf{y}} \in \hat{\mathcal{Y}}_k} \ d_{\text{mces}}(\hat{\mathbf{y}}, \mathbf{y}) \right] \, .
\end{equation*}

\paragraph{Levenshtein distance} Quantifies the minimum number of single-character edits (insertions, deletions, substitutions) required to transform a string $s$ into a string $t$. Given that we denote this distance as $d_{\text{Lev}}(s,t)$, then the corresponding metric is calculated as:
\begin{equation*}
\text{LevDist} = \mathbb{E}_{(\mathcal{S}, \mathbf{y}) \sim \mathcal{D}} \left[ \min_{\hat{\mathbf{y}} \in \hat{\mathcal{Y}}_k} d_{\text{Lev}}(\hat{\mathbf{y}}, \mathbf{y}) \right].
\end{equation*}

\paragraph{Fingerprint-based similarity} To capture substructural similarity, we compute similarity scores using different molecular fingerprints. Let \(f_{\mathrm{fp}}(\mathbf{y})\) be a fingerprint vector of type \(\mathrm{fp} \in \{\text{Morgan}, \text{MACCS}, \text{RDKit}\}\). For each type, the similarity is:
\begin{equation*}
\text{Sim}_{\mathrm{fp}} = \mathbb{E}_{(\mathcal{S}, \mathbf{y}) \sim \mathcal{D}} \left[ \max_{\hat{\mathbf{y}} \in \hat{\mathcal{Y}}_k} \ \text{Tanimoto}\left(f_{\mathrm{fp}}(\hat{\mathbf{y}}), f_{\mathrm{fp}}(\mathbf{y})\right) \right],
\end{equation*}

where the Tanimoto coefficient is:
\begin{equation*}
\text{Tanimoto}(a,b) = \frac{a \cdot b}{\|a\|_1 + \|b\|_1 - a \cdot b} \,.
\end{equation*}
We utilize Morgan, MACCS, and RDKit fingerprints. 
All fingerprints are represented as binary vectors, where each bit indicates the presence (1) or absence (0) of a given feature.
In this setting, the $\ell_1$-norm $\|\cdot\|_1$ corresponds to the number of active bits in the fingerprint. 
\subsection{Failure modes}\label{app:fail_modes}
\begin{figure}[t]
    \centering
    \includegraphics[width=0.98\columnwidth]{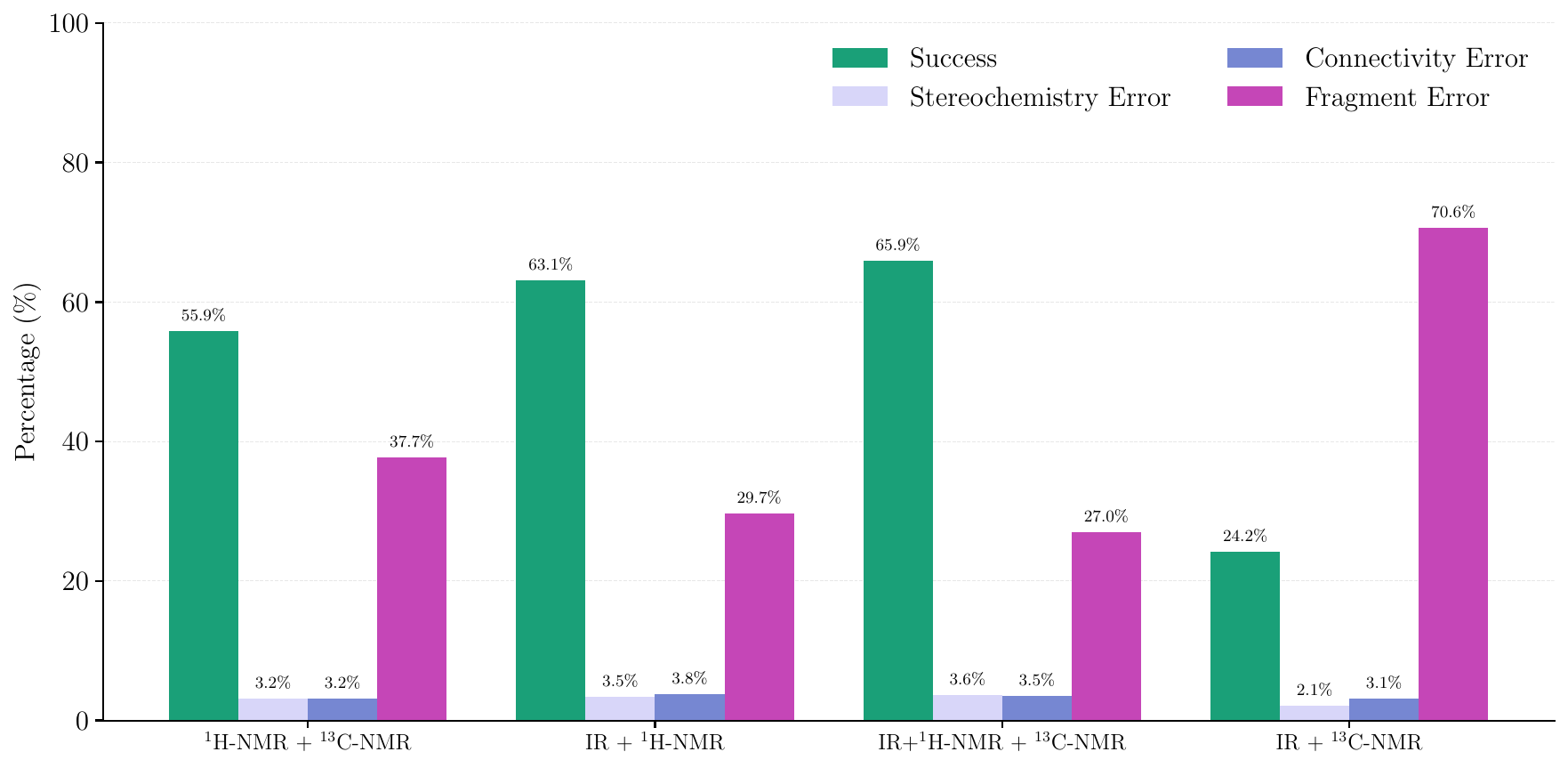}
    \caption{
    Analysis of model failures across error categories for different spectral combinations, alongside the corresponding success rates. Errors are dominated by incorrect fragment predictions, highlighting the need for improved fragment-level representations and tighter spectra–structure alignment.
    }
    \label{fig:failure_modes}
\end{figure}

To better characterize the limitations of the proposed framework, we conduct a systematic analysis of failure cases, summarized in Figure~\ref{fig:failure_modes} and grouped into three main categories: (i) \textit{stereochemistry error} where the model correctly generates the constitutional isomer but fails to reproduce the correct stereochemical configuration of the ground truth molecule, as measured under the enantiomer-aware evaluation protocol; (ii) \textit{connectivity error} in which all relevant functional groups are correctly predicted according to the fragment vocabulary $\mathcal{V}$, but the underlying atomic connectivity is incorrect; (iii) \textit{fragment errors}, where the model fails to predict the correct functional groups or produces spurious fragments not present in the ground truth molecule. 
We observe that the dominant source of failure arises from the misprediction of fragment compositions, suggesting that the model struggles to infer the correct functional group from ambiguous or overlapping spectral evidence. This limitation may stem from the intrinsic ambiguity of spectra-to-molecule mapping or from limited granularity in the fragment vocabulary. While expanding the vocabulary could increase representational expressiveness, it may also amplify combinatorial complexity and learning difficulty. Future work will explore adaptive or hierarchical fragment vocabularies and uncertainty-aware modeling to mitigate these challenges. Moreover, we observe that recent explorations in stereochemistry-aware molecular generation~\citep{tom2025stereo} could help reduce the stereochemistry-related errors observed in our analysis.
\subsection{Fragment occurrences prediction}\label{app:frag_pred}
\begin{figure}[!t]
    \centering
    \includegraphics[width=0.98\textwidth]{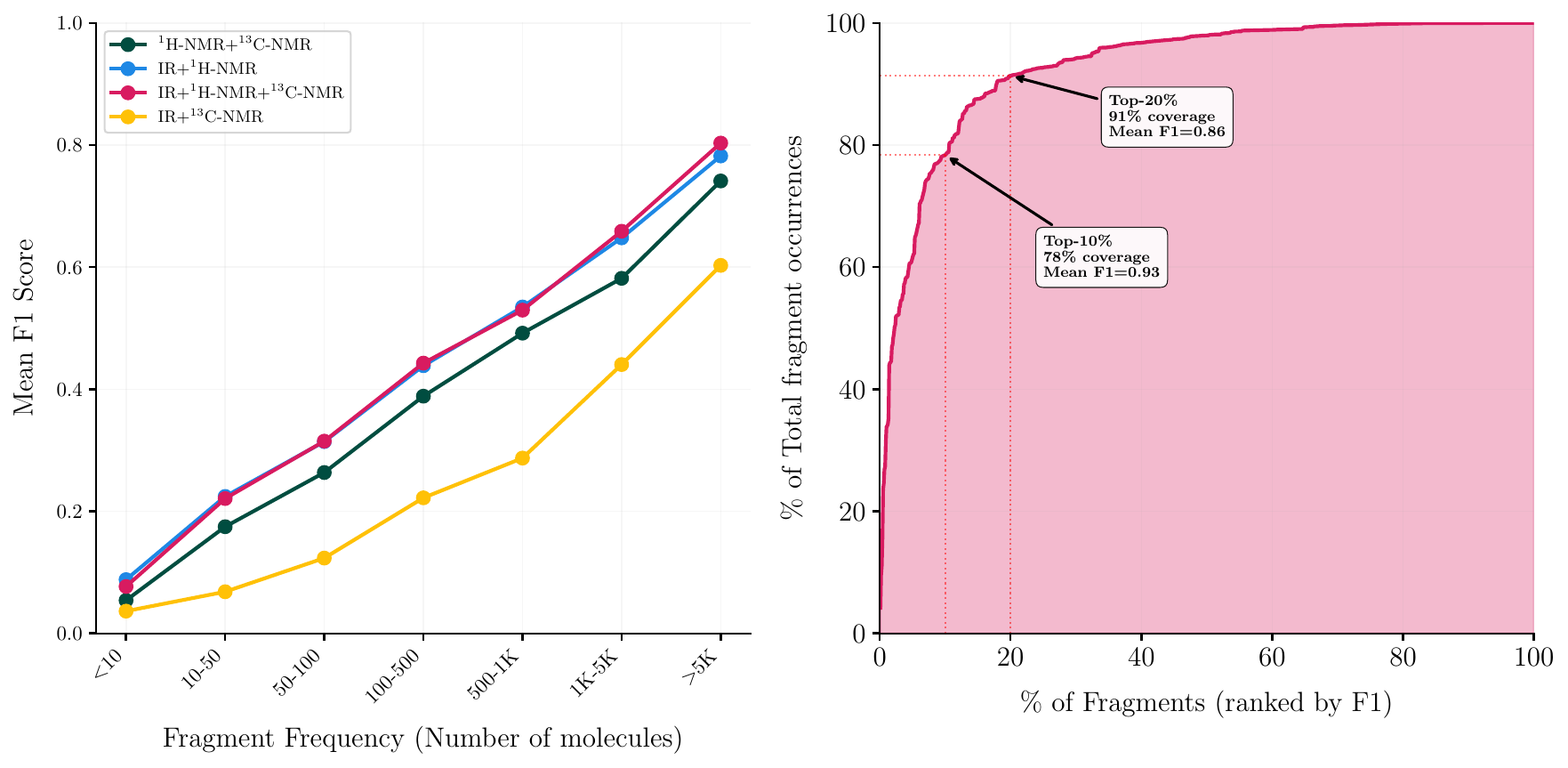}
    \caption{
        (\textit{Left}) Mean F1-score across fragments grouped by molecular frequency. 
        (\textit{Right}) Percentage of all fragment instances in the dataset covered by the top-$X\%$ of fragments ranked by mean F1-score (IR + ${}^{1}$H-NMR + ${}^{13}$C-NMR), showing that high-scoring fragments account for most observed cases.
    }
    \label{fig:fragments_analysis}
\end{figure}

We evaluate fragment predictions using macro- and micro-averaged metrics (Table~\ref{tab:fragment_metrics}). 
Macro metrics weight all fragments equally, while micro metrics aggregate over all predictions and therefore reflect fragment frequency.
We report two complementary criteria: count accuracy, which measures how well the model estimates fragment occurrences, and presence-based metrics (precision, recall, F1), which assess detection irrespective of count. All models achieve high count accuracy ($\approx 0.97$–$0.98$), though this metric is dominated by absent fragments ($c_j=0$) and is less informative under strong class imbalance. 
Presence-based metrics provide a more discriminative assessment. The full multi-spectral model (IR + ${}^{1}$H-NMR + ${}^{13}$C-NMR) achieves the strongest performance, with a micro-averaged F1 of $0.87$ (precision $0.92$, recall $0.82$), indicating that complementary spectral modalities substantially improve fragment identification. In contrast, macro-averaged F1 scores are notably lower (e.g., $0.43$ for the full model), reflecting the difficulty of predicting rare fragments.
Figure~\ref{fig:fragments_analysis} further illustrates these trends. 
Mean fragment-level F1 increases with fragment prevalence, indicating that more common fragments are predicted more accurately.
Complementarily, a cumulative coverage analysis that ranks fragments by their per-fragment F1 shows that the top $10\%$ of fragments account for approximately $78\%$ of all fragment occurrences. This demonstrates that the fragments predicted most reliably by the model also correspond to the most prevalent motifs in the dataset. Consequently, strong performance is concentrated on chemically common fragments, while lower macro-averaged scores primarily reflect the intrinsic difficulty of accurately predicting rare fragments under severe class imbalance. 
Addressing this limitation will likely require targeted data curation or rebalancing strategies to improve generalization across the full fragment vocabulary.
\setlength{\tabcolsep}{1pt} 
\begin{table*}[t]
\caption{Fragment count prediction across spectral combinations.
Micro- and macro-averaged classification metrics.}
\smallskip
\label{tab:fragment_metrics}
\centering
\begin{small}
\begin{sc}
\begin{tabular}{lcccccccc}
\toprule
& \multicolumn{4}{c}{\textbf{Micro-averaged}} 
& \multicolumn{4}{c}{\textbf{Macro-averaged}} \\
\cmidrule(lr){2-5} \cmidrule(lr){6-9}
\textbf{Spectra}
& Accuracy & Precision & Recall & F1
& Accuracy & Precision & Recall & F1 \\
\midrule
$^{1}$H + $^{13}$C-NMR      
& $0.98$ & $0.91$ & $0.78$ & $0.84$
& $0.98$ & $0.60$ & $0.32$ & $0.38$ \\

IR + $^{1}$H-NMR            
& $0.98$ & $0.90$ & $0.83$ & $0.86$
& $0.98$ & $0.60$ & $0.37$ & $0.43$ \\

IR + $^{1}$H + $^{13}$C-NMR 
& $0.98$ & $0.92$ & $0.82$ & $0.87$
& $0.98$ & $0.63$ & $0.37$ & $0.43$ \\

IR + $^{13}$C-NMR           
& $0.97$ & $0.90$ & $0.70$ & $0.79$
& $0.97$ & $0.54$ & $0.20$ & $0.26$ \\
\bottomrule
\end{tabular}
\end{sc}
\end{small}
\end{table*}
\subsection{Outlook and future directions}\label{app:outlook}
This section discusses the primary limitations encountered within the proposed framework and situates them within broader challenges in spectra-to-molecule learning, thereby outlining key directions for future research.
\paragraph{Scale of molecular pre-training}
The fragment-to-molecule pre-training currently leverages a corpus of $\sim3.7$M SMILES, which offers a solid foundation but remains modest compared to the scale of contemporary chemical databases such as PubChem ($\sim$100M molecules).
Given the observed scaling trends in related molecular generative frameworks~\citep{bohde2025diffms}, extending pre-training to larger, more chemically diverse datasets could further enrich the learned molecular prior $p_\phi$, enhancing both fragment composition modeling and downstream spectral elucidation. Future work will investigate large-scale, fragment-conditioned pre-training to assess potential performance gains and improved generalization to rare or complex structures.

\paragraph{2D vs 3D molecular representations}
In this work, we focus on the generation of 2D molecular structures, i.e., atomic connectivity and bond types, rather than explicit 3D conformations. While recent approaches aim to jointly infer 2D and 3D structures from spectra~\citep{wang2025diffspectra}, 2D representations remain a central abstraction for molecular structure elucidation. In particular, 2D graphs capture the core chemical composition and topology of molecules, are directly comparable to existing benchmark datasets, and allow for efficient and scalable generative modeling. Moreover, many spectroscopic modalities provide strong constraints on functional groups and connectivity, even when 3D conformations are ambiguous or undetermined. Incorporating explicit 3D geometry and conformational modeling is an important direction for future work, but lies outside the scope of the present study.

\paragraph{Choice of the prior $p_\phi$}
In this work, the molecular prior $p_\phi$ is learned by reconstructing molecular structures from a coarse, fragment-based representation that encodes both fragment identities and their occurrences. This extends previous binary fragment formulations~\citep{bohde2025diffms, hu2024accurate} by introducing a count-aware model that better reflects the underlying molecular topology.
However, the question of what constitutes an \textit{optimal} prior for downstream spectra-to-molecule task remains open.
An interesting research direction is to study how the design of $p_\phi$ (e.g. by incorporating additional relational structure such as fragment connectivity, local bonding patterns, or hierarchical composition) affects the learnability and transferability of this mapping. In other words, future work should explore priors that are not only chemically-faithful but also \textit{spectroscopically-aligned}, facilitating transferability to the spectra-to-molecule stage.

\paragraph{Simulated vs experimental spectra}
All results in this work are based on simulated spectra generated from computational pipelines that approximate experimental conditions. While this provides consistent supervision, real-world spectra are subject to noise, baseline distortions, solvent effects, and instrument-specific artifacts that may introduce significant distribution shifts.
Adapting the model to such data will require domain adaptation strategies or fine-tuning on curated experimental datasets to ensure robustness under practical laboratory conditions. 

\paragraph{Approximation of fragment inference}
In Eq.~\ref{eq:determ_mapping}, the mapping from spectra $\mathcal{S}$ to fragment composition $\mathbf{c}$ is treated deterministically via $\mathbf{z}_\psi(\mathcal{S})$. 
This neglects inherent ambiguity in the inverse mapping from spectra to substructures (multiple molecular configurations may correspond to highly similar spectral signatures). Future work could relax this assumption by introducing stochastic or variational inference, thereby capturing uncertainty over fragment compositions and improving robustness to ambiguous input spectra.
\subsection{Implementation details} 
\paragraph{Pre-training} 
All models requiring a pre-training stage (NMIRacle, NMR2Struct, and Spec2Mol) are trained using a batch size of $1024$ and a weight decay of $1 \times 10^{-5}$. For NMIRacle and NMR2Struct, we use a learning rate of $1 \times 10^{-5}$ and $\beta = (0.9, 0.98)$. For Spec2Mol, we follow the original implementation settings with a learning rate of $1\times10^{-4}$. While NMIRacle and NMR2Struct are pre-trained on fragments-to-molecule generation, Spec2Mol utilizes a SMILES reconstruction (autoencoding) task.
\paragraph{Spectra-to-molecule fine-tuning} For the downstream task, models are trained for a maximum of $300$ epochs using a batch size of $64$ and the same learning rates as in the pre-training stage (with a learning rate of $1 \times 10^{-5}$ used for the Transformer baselines not leveraging a pre-training stage). We utilize early stopping with a patience of $10$ epochs based on the validation loss specific to each architecture: 
\begin{itemize} 
\item \textbf{NMIRacle (Ours) and NMR2Struct:} The multi-task objective combining SMILES reconstruction with fragment-level supervision. Crucially, we monitor the \textit{fragment count} loss for NMIRacle, compared to the \textit{binary presence} classification for NMR2Struct. 
\item \textbf{Spec2Mol:} The weighted combination of $\ell_2$ latent alignment loss and SMILES reconstruction loss. 
\item \textbf{SMILES/SELFIES Transformers:} The standard sequence cross-entropy loss. 
\end{itemize} For generation, all models utilize top-$k$ sampling with $k=5$ and temperature $T=1.0$.
\newpage
\subsection{Architecture details}
In Table~\ref{tab:architecture} we present a summary of the main architectural components of NMIRacle. 
\begin{table}[!h]
\centering
\setlength{\tabcolsep}{22pt} 
\caption{NMiracle Model Architectural Details}
\vspace{1mm}
\label{tab:nmiracle_architecture}
\begin{tabular}{llc}
\toprule
\textbf{Component} & \textbf{Parameter} & \textbf{Value} \\
\midrule
\multirow{1}{*}{\textbf{Global}} 
& Model Dimension ($d$) & 128 \\
\midrule
\multirow{3}{*}{\textbf{Fragment Encoder}} 
& Vocabulary Size & 991 \\
& Embedding Dimension & 128 \\
& Activation & GELU \\
\midrule
\multirow{6}{*}{\textbf{Transformer Model ($p_\phi$)}} 
& Encoder Layers & 6 \\
& Decoder Layers & 6 \\
& Attention Heads & 8 \\
& FFN Dimension & 1024 \\
& Dropout & 0.1 \\
& Activation & ReLU \\
\midrule
\multirow{10}{*}{\textbf{Multispectra Encoder ($q_\psi$)}} 
& Kernel Size 1 & 5 \\
& Pool Size 1 & 12 \\
& Out Channels 1 & 64 \\
& Kernel Size 2 & 9 \\
& Pool Size 2 & 20 \\
& Out Channels 2 & 128 \\
& Transformer encoder layers & 2 \\
& Attention Heads & 4 \\
& Activation & ReLU \\
& $^{13}$C-NMR binary bins & 80 \\
\midrule
\multirow{2}{*}{\textbf{Fragment composition head}}
& Hidden dimension & 256 \\
& Activation & GELU \\
\bottomrule
\end{tabular}
\label{tab:architecture}
\end{table}
\newpage
\subsection{Success cases}\label{app:success_cases}
\begin{figure}[!h]
    \centering
    \includegraphics[width=0.98\textwidth]{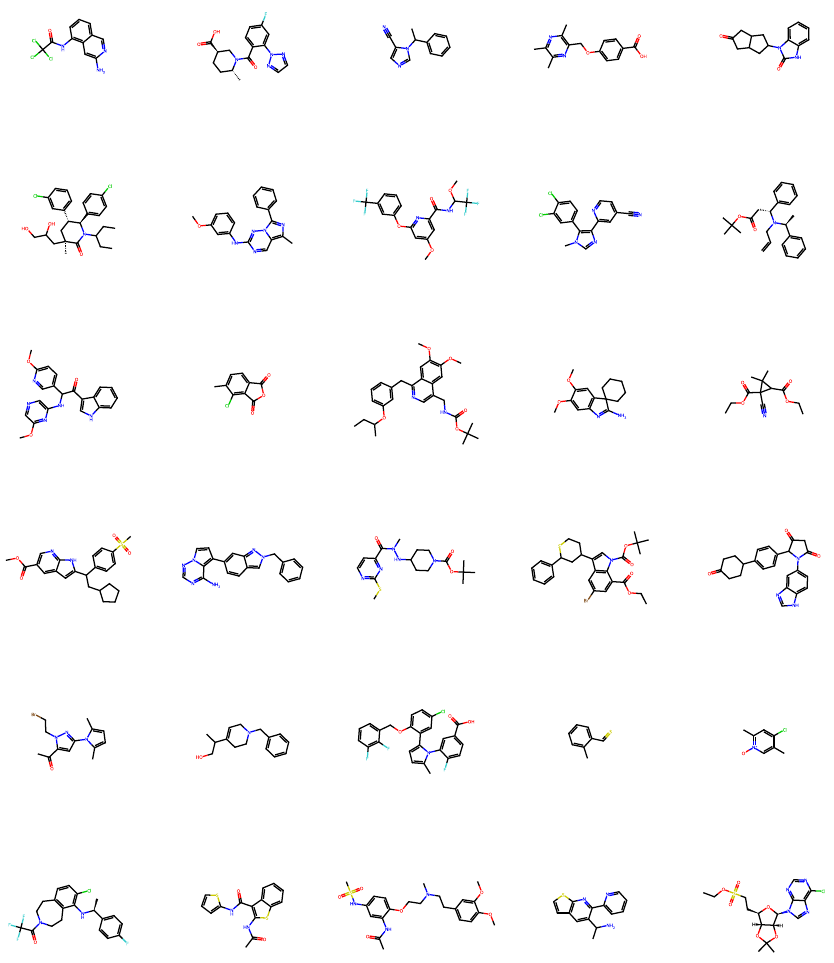}
\caption{
Representative success cases in which NMIRacle correctly predicts the ground-truth molecule, while all baseline models fail.
For each example, the prediction from NMIRacle matches the ground truth structure under the enantiomer-aware evaluation protocol, whereas competing methods do not recover the correct molecular structure among their top-$k$ candidates.
}
\label{fig:success_cases}
\end{figure}

Figure~\ref{fig:success_cases} illustrates representative success cases under full multi-spectral conditioning (IR, ${}^{1}$H-NMR, ${}^{13}$C-NMR), where NMIRacle correctly recovers the target molecular structure, while all competing methods fail under the same enantiomer-aware evaluation protocol.

\end{document}